\documentclass[prb,twocolumn,amsmath,amssymb,floatfix]{revtex4}
\usepackage{graphicx}% Include figure files
\usepackage{bm}% bold math

\newcommand{\be}{\begin{equation}}
\newcommand{\ee}{\end{equation}}
\newcommand{\beq}{\begin{eqnarray}}
\newcommand{\eeq}{\end{eqnarray}}

\begin{document}

\def \tr{{\mbox{tr~}}}
\def \ra{{\rightarrow}}
\def \ua{{\uparrow}}
\def \da{{\downarrow}}
\def \be{\begin{equation}}
\def \ee{\end{equation}}
\def \ba{\begin{array}}
\def \ea{\end{array}}
\def \bea{\begin{eqnarray}}
\def \eea{\end{eqnarray}}
\def \nn{\nonumber}
\def \l{\left}
\def \r{\right}
\def \half{{1\over 2}}
\def \etal{{\it {et al}}}
\def \cH{{\cal{H}}}
\def \cM{{\cal{M}}}
\def \cN{{\cal{N}}}
\def \cQ{{\cal Q}}
\def \cI{{\cal I}}
\def \cV{{\cal V}}
\def \cG{{\cal G}}
\def \bS{{\bf S}}
\def \bI{{\bf I}}
\def \bL{{\bf L}}
\def \bG{{\bf G}}
\def \bQ{{\bf Q}}
\def \bR{{\bf R}}
\def \br{{\bf r}}
\def \bu{{\bf u}}
\def \bq{{\bf q}}
\def \bk{{\bf k}}
\def \bz{{\bf z}}
\def \bx{{\bf x}}
\def \tJ{{\tilde{J}}}
\def \W{{\Omega}}
\def \e{{\epsilon}}
\def \lam{{\lambda}}
\def \L{{\Lambda}}
\def \a{{\alpha}}
\def \t{{\theta}}
\def \b{{\beta}}
\def \g{{\gamma}}
\def \D{{\Delta}}
\def \d{{\delta}}
\def \w{{\omega}}
\def \s{{\sigma}}
\def \f{{\varphi}}
\def \x{{\chi}}
\def \e{{\epsilon}}
\def \h{{\eta}}
\def \G{{\Gamma}}
\def \z{{\zeta}}
\def \hatt{{\hat{\t}}}
\def \hn{{\bar{n}}}
\def \vk{{\bf{k}}}
\def \vq{{\bf{q}}}
\def \gk{{\g_{\vk}}}
\def \nd{{^{\vphantom{\dagger}}}}
\def \yd{^\dagger}
\def \av#1{{\langle#1\rangle}}
\def \ket#1{{\,|\,#1\,\rangle\,}}
\def \bra#1{{\,\langle\,#1\,|\,}}
\def \braket#1#2{{\,\langle\,#1\,|\,#2\,\rangle\,}}

\title{Decay of a superfluid currents
in a moving system of strongly interacting bosons}

\author{A.~Polkovnikov, E.~Altman, E.~Demler, B.~Halperin and M.D.~Lukin}
\address{Physics Department, Harvard University, Cambridge, MA 02138}
\date{\today}

\begin{abstract}
We analyze the stability and decay of supercurrents of strongly
interacting bosons on optical lattices. At the mean field level,
the system undergoes an irreversible dynamic phase transition,
whereby the current decays beyond a critical phase gradient that
depends on the interaction strength. At commensurate filling the
transition line smoothly interpolates between the classical
modulational instability of weakly interacting bosons and the
equilibrium Mott transition at zero current. Below the mean field
instability, the current can decay due to quantum and thermal
phase slips. We derive asymptotic expressions of the decay rate
near the critical current. In a three dimensional optical lattice
this leads to very weak broadening of the transition. In one and
two dimensions the broadening leads to significant current decay
well below the mean field critical current. We show that the
temperature scale below which quantum phase slips dominate the
decay of supercurrents, is easily within experimental reach.
\end{abstract}

\maketitle
\section{Introduction}

Many-body physics of strongly interacting ultra-cold atoms in
optical lattices has been actively explored in the recent years.
In particular, quantum effects of strongly interacting bosons,
such as number squeezed states generation~\cite{orzel} and a
quantum phase transition from the superfluid to the Mott
insulator~\cite{bloch}, have been observed in agreement with
earlier theoretical predictions~\cite{fisher, jaksch}. These
developments were followed by a broader theoretical analysis of
phase diagrams of more complex systems including multi component
bosons\cite{zhou,kuklov,kuklov1,ehud1, cazalilla}, Bose-Fermi
mixtures~\cite{charles,pazy,buchler}, and exotic states exhibiting
topological orders~\cite{zhou,lesik}. Such studies are motivated
by issues that arise in a traditional condensed matter context.
However, unique features of the cold atom systems also raise a
completely new set of questions. In particular, the ability to
continuously vary interaction parameters, coupled to the near
perfect isolation of these systems, open the way to address
quantum dynamics far from equilibrium.

In this context, there exists a purely dynamical phase transition
of a condensate of weakly interacting bosons moving in an optical
lattice. If the wave vector associated with the condensate
momentum exceeds a critical value, which is equal to one quarter
of the reciprocal lattice constant for the square
lattice~\cite{niu,bishop}, then the coherent motion of the
condensate becomes unstable, resulting in the loss of
superfluidity. Such a dynamical instability was observed
experimentally~\cite{inguscio} by measuring loss of coherence as a
function of the condensate momentum. This transition is of
classical origin, in the sense that it is seen as an instability
in the Gross-Pitaveskii equations of motion (GPE). Reltaed
nonlinear dynamical phenomena such as self trapping and soliton
formation have been studied in theory and experiment
\cite{smerzi1,oberthaler,oberthaler1, brazhnyi1}. However, these
studies focused on essentially classical systems, well described
by GPE. Very little progress has been made in analyzing far from
equilibrium behavior of systems where strong interactions and
quantum fluctuations play an important role.

In the present work we address this issue by focusing on a problem
relevant to recent experiments: the fate of superfluid currents in
optical lattices in the strongly interacting regime. We shall
consider this issue via two questions, which will turn out to be
closely related. First, what is the effect of quantum (as well as
thermal) fluctuations on the dynamical instability of a moving
condensate? May the instabilitiy occur earlier, for example, in
this case than the GPE prediction of $p_c=\pi/2$? The second
question is, how does the superfluid to Mott insulator transition
take place when the condensate is moving in the lattice. This
question may have direct relevance for ongoing experiments on the
superfluid-insulator transition. When the condensate is loaded
into a magnetic trap it is hard to completely avoid center of mass
oscillations. In the absence of the optical lattice such a motion
is frictionless and can persist for very long times. The same is
true in the superfluid phase in the presence of the optical
lattice as long as the center of mass momentum remains small and
the interactions are weak.
But what is the ultimate fate of this motion as
the optical potential is increased and the system approaches the
insulating regime?

Effect of weak quantum fluctuations on the modulational
instability in one dimensional traps was analyzed earlier
numerically by one of us~\cite{pw}. It was shown that the quantum
fluctuations smoothen the sharp classical transition and lead to
the current decay at smaller amplitude of the center of mass
oscillations than predicted using the classical
Gross-Pitaevskii (GP) equations alone. Similar numerical results
were also reported in Ref.~[\onlinecite{nist}]. Recent experiments
confirmed strong damping of the center of mass oscillations in one
dimensional condensates far from the classical modulational
instability~\cite{chad}.

In this paper we present a general theoretical framework of a
superfluid-insulator transition in the current carrying state.
Strictly speaking this is a true phase transition only at zero
current. However, we find regimes where the broadening
of the transition is small and even where a true discontinuity
survives. In such cases a phase boundary is still well-defined.

We shall show that at any nonzero current the transition is
irreversible. If one starts from a condensate with nonzero
current, increases the lattice strength past the transition point,
then decreases it back to the original state, the current in the
final state will have vanished. The energy contained in the
initial motion of the condensate is transferred into thermal
incoherent excitations. Thus the dynamical transition is of the
first order type.  The current carrying state is a metastable
minimum of the classical (saddle-point) energy, and the transition
occurs when the system escapes from this state.

Throughout this work we employ the well known boson
Hubbard model~\cite{sachdev_book}, described by the Hamiltonian:
\be
\mathcal H=\sum_{<jk>} -J(a_j^\dagger a_k+a_k^\dagger
a_j)+\sum_j{U\over 2}a_j^\dagger a_j(a_j^\dagger a_j-1),
\label{bh}
\ee
where $a_j^\dagger$ and $a_j$ are the boson creation and
annihilation operators at the lattice site $j$, $<jk>$ denotes
pairs of nearest neighbors, $J$ is the single particle hopping
amplitude and $U$ is the on-site repulsive interaction. Another
implicit parameter is the average number of bosons per site $N$.
If also a condition $UN\gg J$ is fulfilled, then the boson Hubbard
model can be mapped into the $O(2)$ quantum rotor
model~\cite{sachdev_book}:
\be
\mathcal H=\sum_{<jk>} -2JN\cos(\phi_{k}-\phi_j)+\sum_j{U\over
2}n_j^2,
\label{hamilt}
\ee
where $\phi_j$ and $n_j$ are the conjugate phase and the number of
particles on the site $j$. The system described by (\ref{hamilt})
also undergoes a superfluid insulator transition at $JN\sim U$ and
can support current in the superfluid phase. In many situations,
the quantum rotor model is easier to analyze analytically and we
will frequently appeal to it. It is usually well justified in the
case $N\gg 1$. Indeed, in this limit it is possible to have
simultaneously $UN\gg J$ and $JN$ either smaller or larger than
$U$. So at large $N$ the mapping from the boson Hubbard to the quantum
rotor model can be justified in both the superfluid and insulating
phases.

Our paper is organized as follows. In sec.~\ref{sec0} we give an
overview of our main results and present a physical picture. In
sec.~\ref{sec1} we derive the mean field phase diagram separating
stable and unstable regimes of current flow. Sec.~\ref{sec2}
focuses on the current decay mechanisms due to quantum and thermal
fluctuations. In particular, we obtain leading asymptotic
contributions to both quantum and thermal decay rates near the
mean field instability. Then in sec.~\ref{sec3} we consider
dynamics of the current decay and discuss the effects of a
parabolic confining potential. Sec.~\ref{sec4} addresses the loss
of coherence in a condensate following  super-current decay. In
sec.~\ref{sec:exact} we present the results of exact simulations
in small systems and discuss them in the context of our
theoretical analysis. Finally, in sec.~\ref{sec:fin} we summarize
the results and discuss experimental implications. A shorter
discussion of the results presented here can be found
in~Ref.~[\onlinecite{prl}].

\section{Physical picture and overview of the results.}
\label{sec0}

The existence of a critical velocity of a condensate in an optical
lattice was predicted~\cite{niu, bishop} and observed
experimentally ~\cite{inguscio,oberthaler} in the regime of weak
quantum fluctuations ($JN\gg U$). In this case one can solve the
Gross-Pitaevskii equations of motion and find that the condensate
becomes unstable when the phase difference between adjacent sites
becomes larger than $\pi/2$. There is a simple way to understand
this instability by considering how the superfluid current flowing
through the system changes with the condensate momentum. In a
continuum system the current is just
\be
\mathcal J=\rho_s p,
\ee
where $\rho_s$ denotes the superfluid density and $p$ the momentum
or phase gradient such that $\phi(x)=px$. In a discrete system
described by a tight-binding model, the above expression is
modified to
\be
\mathcal J=\rho_s \sin p.
\label{cur}
\ee
More generally $\sin p$ is replaced by a different function of
$p$, with the same elementary period.
At small currents  we recover the continuum limit from
Eq.~(\ref{cur}). In the
Gross-Pitaevskii regime, the superfluid density does not depend on
momentum. Therefore the maximal current occurs at
$p=\pi/2$, precisely where the condensate motion becomes unstable.
As we argue below, this is not a coincidence.
\begin{figure}[ht]
\includegraphics[width=8cm]{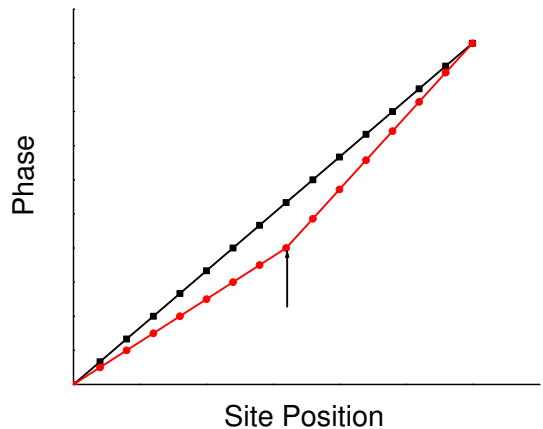}
\caption{Schematic representation of a perturbation to a state
with a uniform phase gradient. Dots represent phases on different
sites for uniform and perturbed systems. The lines are guides to
an eye.}
\label{fig_kink}
\end{figure}
Consider a perturbation, where the state with a uniform phase
gradient $p$ is split into two equal domains with slightly higher
and slightly lower momenta $p\pm\delta p$ (see
Fig.~\ref{fig_kink}). At small $\delta p$ we can expand the energy
difference between the perturbed and unperturbed configurations in
powers of $\delta p$. The linear term in the expansion vanishes
because the contributions to the energy from the left and the
right domains exactly cancel each other and we are left only with
the quadratic term:
\be
\delta E\approx {1\over 2}{dE\over dp}\delta p+{1\over 2}{dE\over
dp}(-\delta p)+2 {1\over 4} {d^2 E\over dp^2}(\delta p)^2={1\over
2} {d^2 E\over dp^2}(\delta p)^2,
\ee
Noting that the superfluid current is formally defined as the
derivative of the energy with respect to phase gradient: $\mathcal
J(p)=dE/dp$ we find
\be
\delta E\propto {d\mathcal J\over d p}(\delta p)^2.
\ee
Thus if the current is an increasing function of the momentum,
then small deviations from the uniform state increase its energy.
On the other hand if $d\mathcal J/dp<0$ then the fluctuation just
considered reduces the energy of the system. In this case it is
obvious there is a manifold of resonant configurations obtained by
a smooth continuum transformation from a uniform state. For
example one can take the state in Fig.~\ref{fig_kink} plus a weak
positive energy fluctuation. So there is no energy barrier
protecting a uniform state from fragmentation. Let us note that
this argument shows that the positive slope of the current with
respect to $p$ is a necessary condition for the stability of the
uniform state. It does not exclude, however, that the current can
decay even if this condition is satisfied.
In sec.~\ref{sec2} we will see this indeed occurs
due to quantum or thermal phase slips~\cite{tinkham2,zaikin}.

Consider now the strongly interacting regime.
Suppose that we deal with a uniform system at commensurate (i.e.
integer) filling. It is well known that such systems undergo a
superfluid-insulator transition~\cite{fisher} at zero temperature.
This transition is driven by  quantum fluctuations which increase
with the interaction strength. As $p$ increases the effective
hopping amplitude in the direction parallel to the current
decreases as $J_{eff}\to J\cos p$ resulting in  reduction of the
sound velocity~\cite{bishop}. Alternatively reduction of $J_{eff}$
can be viewed as increase of the single-particle effective mass
with quasi-momentum. This immediately follows from a
single-particle band structure. As a result, quantum fluctuations,
which are determined by the ratio
$(U/J_{eff}N)$~[\onlinecite{psg}], become stronger with $p$,
implying concomitant increase of quantum depletion of the
superfluid density. Therefore the equation (\ref{cur}) should be
rewritten as:
\be
\mathcal J=\rho(p)\sin p,
\label{cur1}
\ee
where $\rho(p)$ is a monotonically decreasing function of the
momentum. Thus the product reaches a maximum at some
$p^\star<\pi/2$. In the Gross-Pitaevskii regime $JN\gg U$, the
dependence of $\rho(p)$ on $p$ is very weak and we find that
$p^\star\approx \pi/2$. On the other hand, in the vicinity of the
superfluid-insulator transition $\rho(p)$ is both very small and
very sensitive to variations of $J_{eff}$. Thus we expect that in
this case $p^\star$ will be close to zero.

In sec~\ref{sec1} we give a formal derivation of the critical
momentum at which the condensate motion becomes unstable. Using
the time dependent Gutzwiller approximation,  we plot the critical
momentum as a function of the interaction strength. This phase
boundary interpolates between the usual dynamical instability
occuring at $p=\pi/2$ for small interactions and the vanishing
critical momentum at the equilibrium superfluid-insulator
transition (see the top graph in Fig.~\ref{ph-dg2}).

The situation is different in the non-commensurate case . No
matter how strong the interaction strength, it can not localize
the excess particles (or holes)     moving on top of the Mott
background. This excess density always remains superfluid,
independent of $J_{eff}$ and thus also of $p$. Together with
Eq.~(\ref{cur1}), this implies that at strong interactions the
instability occurs at $p=\pi/2$. On the other hand for
sufficiently weak interactions, where the number fluctuation per
site $\d N>>1$, there is no distinction between integer and non
integer density. Therefore for $U$ not too large the critical
momentum should decrease with $U$ from $\pi/2$. Indeed we find,
using the time dependant Gutzwiller approximation, that for the
incommensurate case $p_c$ reaches a minimum at some finite
interaction strength and saturates on $\pi/2$ for both very weak
and very strong interactions. (see the bottom graph in
Fig.~\ref{ph-dg2}).

In Sec.~\ref{sec3b} we develop an analytical approach, which
describes the behavior of the critical momentum $p_c$ versus
interaction in the vicinity of the zero-current SF-IN transition.
We show that in the commensurate case $p_c$ vanishes as $1/\xi$,
where $\xi$ is a coherence length, which diverges at the
transition. At non-integer filling we confirm the non monotonic
behavior of the critical momentum.

In practice the system always includes a global harmonic
confinement, which leads to a non uniform density distribution. In
this case we find that the instability first occurs at the the
edges of the condensate where $N=1$ regardless of the peak density
$N_0$ in the middle of the trap (see Sec.~\ref{sec3}). There is a
difference between large and small $N_0$, which manifests in the
dynamics of the current decay. For $N_0\approx 1$ (as well as in
uniform systems with arbitrary filling) we find that near the
instability the decay is underdamped, i.e. the instability rapidly
grows in time destroying the current state. On the other hand if
$N_0$ is large, the momentum oscillations decay gradually after
the edges become unstable.  There is another important difference
between the uniform and parabolically trapped condensates. In a
uniform lattice there are only two energy scales, set by $U$ and
$J$ and their ratio completely determines the behavior of the
system. With harmonic confinement on the other hand, due to the
presence of another energy scale one should take into
consideration whether $U$ or $J$ or both are being changed in the
experiment (see discussion in Sec.~\ref{sec3} and
Appendix~\ref{seca3}).

\begin{figure}[ht]
\includegraphics[width=8cm]{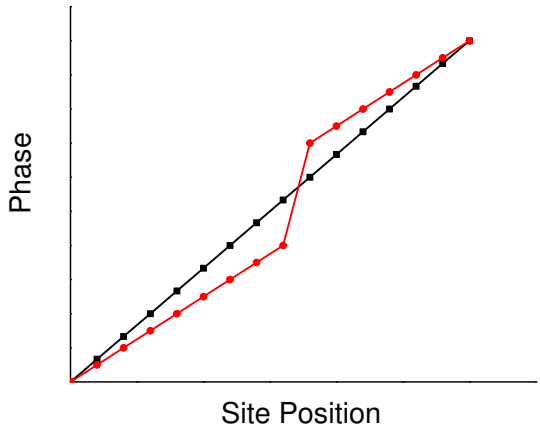}
\caption{Schematic representation of a phase-slip. Notations are
the same as in Fig.~\ref{fig_kink}.}
\label{fig_slip}
\end{figure}

So far we concentrated on the results of the mean field dynamics
at zero temperature, where the time evolution is simply described
by classical equations of motion. In Sec.~\ref{sec2} we go beyond
the mean field dynamics to analyze the effect of quantum and
thermal fluctuations. These act to generate phase-slips, which
induce current decay even prior to the classical instability. A
phase configuration for a particular phase slip is shown in
Fig.~\ref{fig_slip}.
Basically a phase-slip corresponds to generation of a large phase
difference on a particular link (or in the vicinity of this link)
and simultaneous reduction of the phase gradient in the rest of
the chain. Because the energy is a periodic function of phase
differences, by generating phase slips the system reduces its
superfluid current. We calculate the leading exponents of the
decay rates in the Gross-Pitaevskii regime of relatively weak
interactions and in the quantum critical regime close to the SF-IN
transition. We find, that broadening of the mean field transition
is strongest in the one-dimensional case. In particular, deep in
the superfluid regime, the phase-slip tunneling rate at
$p\to\pi/2$ scales as
\be
\Gamma\propto \exp\left(-S_d\sqrt{JN\over U}(\pi/2-p)^{6-d\over
2}\right),
\ee
where $S_d$ is just a number that depends on the number of space dimensions
(see Eqs.~(\ref{s_1d})-(\ref{s_3d})). We
obtain similar results for the thermal decay. Although below
$\pi/2$ the decay in 3D is clearly much weaker than in 1D, there
is no qualitative difference between various dimensions. As long
as the ratio $JN/U$ remains large, the tunneling of phase-slips is
exponentially suppressed.
%Note that it is also necessary that
%$UN/J$ remains of the order of unity or larger in order to have a
%stable superfluid at large momentum $p\sim\pi/2$. Otherwise, even
%at the classical Gross-Pitaveskii level, the condensate can be
%easily destroyed due to Landau instability at small
%$p$~[\onlinecite{niu}]. The Landau instability is not specific to
%the lattice systems and can occur in uniform condensates, when the
%condensate velocity exceeds the sound velocity
%and provided there is some breaking of translational symmetry.
%However, in the strongly interacting regime $UN\gg J$ both
%instabilities coincide at $p=\pi/2$~\cite{niu} and thus there is no
%distinction between them. The two conditions $JN\gg
%U$ and $UN\gtrsim J$ can be satisfied simultaneously
%only at large filling $N\gg 1$.
%This is typically the case in experiments with
%quasi-one-dimensional condensates~\cite{inguscio}.

Next we derive analytical expressions for the exponents
characterizing the decay rate in the vicinity of the equilibrium
SF-IN transition. We find again that fluctuation induced decay is
stronger in lower dimensions. However, there is no small parameter
like $U/JN$ controlling the mean field results. We show that in
one dimension the exponent always remains of the order of one,
implying significant broadening of the mean field transition. In
three dimensions, by contrast, we find that the quantum decay rate
does not vanish at the mean field instability, but rather exhibits
a discontinuous jump. In this sense, the three dimensional system
undergoes a sharp dynamical transition at zero temperature.

The physical picture of current decay due to generation of phase
slips is similar to the situation in superconductors. In
particular, deviation of the critical current from the mean field
result was observed  in thin superconducting wires~\cite{webb} and
explained theoretically~\cite{langer1, halperin}. The mechanism
responsible for reduction of the critical current was identified
as creation of phase slips due to thermal fluctuations. The
question of observing current decay in superconductors due to
quantum tunneling is still debated (see
Ref.~[\onlinecite{tinkham2}] for recent developments). We will
show that in the systems under consideration here, current decay
due to quantum effects is easily within experimentally reach. For
a one dimensional system, for example, the characteristic
temperature below which quantum decay dominates in the GP regime
(far from the Mott transition) is of the order of the Josephson
energy ($T^\star\approx \sqrt{UJN}$). This is much higher than
typical temperatures in optical lattice experiments. At strong
interactions, in the vicinity of the Mott transition, broadening
of the mean field transition due to quantum phase slips is always
large in one and two dimensions. Only in the three dimensional
case, where quantum tunneling is suppressed, thermal fluctuations
can be responsible for current decay below the mean field
transition. Some additional details on the relation between
current decay in superconductors and in optical lattices can be
found in Ref.~[\onlinecite{tinkham1}].

Let us now briefly mention some interesting experimental
implications of our results. We envision the following
experimental scheme. Start with a superfluid state far from the
Mott insulator. Then either boost the condensate to some non zero
momentum~\cite{inguscio}, or induce a center of mass oscillation
in the trap~\cite{chad}. Now, slowly increase the interaction
parameter up to a specified point. This can be done in the usual
way by increasing the lattice intensity, or alternatively by
decreasing the detuning from a Feshbach resonance. Finally
decrease the interactions back to their original value. If the
dynamical phase transition is sharp then as long as the system
does not cross the transition boundary (Fig. \ref{ph-dg2}a) within
this cycle, the process should be completely reversible. In
particular, the initial current (or center of mass oscillation),
as well as phase coherence, should be fully recovered. At the same
time, if the system does pass through the transition, the current
will be lost irreversibly and the system will heat and partially
lose its coherence, compared to the original state.

One of our main results, is that in a three dimensional optical
lattice this mean field dynamical transition is sharp, and it
essentially survives the effect of fluctuations. Such experiments
can thus map the non-equilibrium phase diagram shown in (Fig.
\ref{ph-dg2}) and directly trace the connection between the
classical modulational instability ($p_c=\pi/2$) and the the
equilibrium Mott transition. In fact, due to it's discontinuous
nature, the dynamical transition point is much easier to detect.
This suggests an accurate method to determine the position of the
equilibrium Mott transition by extrapolating the dynamical
transition line to zero momentum.

One comment is in order concerning heating and loss of coherence
in the final state. In Sec.~\ref{sec4} we show that in three
dimensions this loss of coherence is only significant at very
large currents ($p\sim \pi/2$). In one dimension (and to a lesser
extent in two dimensions), the loss of coherence due to
irreversible heating depends on the system size or experimental
resolution and may thus be large even at small currents.

\section{Mean field dynamics and critical currents in optical lattices.}

\label{sec1}

\subsection{Gross-Pitaveskii regime}

Despite its simple form, the Bose Hubbard model (\ref{bh}) is not
integrable in any spatial dimension~\cite{shastry,kolovsky} and
can not be solved completely. Nevertheless, there are some limits
where one can make considerable progress in understanding its
static and dynamic properties. In particular, one can easily
address the regime of weak quantum fluctuations, which is the case
when $JN\gg U$~\cite{psg}. Then one can use discrete
Gross-Pitaveskii equations~\cite{stringari}. For the
Hamiltonian~(\ref{bh}) these are given by:
\be
i{d\psi_j\over d t}=-J\sum_{k\in O}\psi_{k}+U|\psi_j|^2\psi_j,
\label{gp1}
\ee
where the classical fields $\psi_j$ and $\psi_j^\star$ correspond
to the expectation values of $a_j$ and $a_j^\dagger$ respectively;
the set $O$ contains the nearest neighbors of site $j$. In the
quantum rotor limit $UN\gg J$ the number fluctuations are weak and
can be integrated out leaving us with equations of
motion for the phase $\phi_j=\arg \psi_j$ only:
\be
{d^2\phi_j\over dt^2}=-2UJN\sum_{k\in O}\sin(\phi_{k}-\phi_j).
\label{gp2}
\ee
Both equations (\ref{gp1}) and (\ref{gp2}) can support stationary
current carrying states $\psi_j\propto \exp(ipj)$. A simple linear
stability analysis shows that these states are
stable towards small perturbations for $p<\pi/2$ and become
unstable otherwise~\cite{niu, bishop}.
Thus, $\pi/2$ is the critical phase twist above
which a uniform superfluid state breaks down. The transition from
the superfluid to the insulating state associated with this
instability is known as the classical localization transition. It
was recently observed experimentally~\cite{inguscio}. In the
presence of quantum
fluctuations, the current can decay even for
$p<\pi/2$ via quantum tunneling. Clearly these fluctuations should
be increasingly important as the system approaches the Mott
phase. The same is true for decay due to thermal fluctuations
as one increases the temperature.
In the next section we will address this issue in
detail.

\subsection{Critical current in the vicinity of the SF-IN transition}
\label{sec3b}

The Gross-Pitaveskii description of the dynamics breaks down at
strong interactions. Moreover when $JN\sim U$ the bosonic system
at commensurate filling ($N$ is integer) undergoes the Mott
insulator transition entirely driven by quantum
fluctuations~\cite{fisher, sachdev_book}. In the uniform system
with a fixed density this transition lies in the universality
class of the $xy$ model in $d+1$ dimensions~\cite{sachdev_book},
the properties of which were well studied long
ago~\cite{Zinn-Justin}. So there is also a hope to get insights to
some nonequilibrium properties of the interacting bosons in the
vicinity of the phase transition. The latter, as any generic
second order phase transition, is characterized by a diverging
correlation length $\xi$~[\onlinecite{sachdev_book}], which sets
the length scale for all low-energy universal properties of the
system. In particular, close to the critical point the low-energy
long-wavelength fluctuations can be described by relativistic
($z=1$) effective field theory~\cite{sachdev_book}. In terms of
dynamics this implies that the classical equations of motion also
take explicitly relativistic invariant form~\cite{ehud}:
\be
{\partial^2\psi\over \partial t^2}={\bf
\nabla}^2\psi+r\psi-|\psi|^2\psi,
\label{gpr}
\ee
where $\psi$ is the superfluid order parameter, $r$ tunes the
system across the SF-IN transition: $r>0$ corresponds to the
superfluid phase and $r<0$ does to the insulator. Here we rescaled
the units of coordinates and time by a constant of the order of
one (see Appendix~\ref{sec:ax} for the details). The correlation
length $\xi$ is related to the tuning parameter $r$ as $\xi\propto
1/\sqrt{|r|}$. We point out that Eq.~(\ref{gpr}) is very
reminiscent of the conventional continuum Gross-Pitaeskii equation
with the only difference that there is a second (as opposed to
first) order time derivative in the LHS. This equation has a
conserved charge:
\be
Q=\int d^dx {A\over
2i}\left(\psi^\star\partial_t\psi-\psi\partial_t\psi^\star\right),
\ee
which is proportional to the deviation of the particle density
from the integer filling; the constant $A$ is explicitly given in
the Appendix~\ref{sec:ax}. Thus the stationary solutions
correspond to the commensurate transition. In the non-commensurate
regime there is no phase transition, but one can still use
Eq.~(\ref{gpr}) if the deviation from the integer filling is
small.

\subsubsection{Commensurate case}

Let us analyze the fate of the current-carrying case if the
mean number of bosons per site is integer.  Equation~(\ref{gpr})
supports stationary states of the form:
\be
\psi(x,{\bf z})=\sqrt{r-p^2}\mathrm e^{ip x},
\label{psi2}
\ee
where ${\bf z}$ denotes $d-1$-dimensional space of transverse
coordinates. Without any further analysis it is obvious that
current states become unstable at $p\gtrsim \sqrt{r}\sim 1/\xi$,
i.e. the critical momentum vanishes at the superfluid-insulator
transition point. To be more precise, we can evaluate the spectrum
of small fluctuations of Eq.~(\ref{gpr}) around the stationary
solution~(\ref{psi2}):
\be
\omega^2({\bf q})=r-p^2+{\bf q}^2\pm
\sqrt{(r-p^2)^2+4p^2q_x^2\xi^4},
\ee
where $q$ is the wavevector characterizing the fluctuations around
the state (\ref{psi2}). In the long wavelength limit $|{\bf q}|\to
0$ the expression above yields simplified frequencies for the
amplitude and the phase modes:
\beq
&&\omega_1^2({\bf q})\approx 2(r-p^2)+{r+p^2\over
r-p^2}q_x^2+{\bf q}_\perp^2,\label{omega1}\\
&&\omega_2^2({\bf q})\approx q_x^2 {r-3p^2\over r-p^2}+{\bf
q}_\perp^2\label{omega2}.
\eeq
The first (amplitude) branch is always gapped unless $p^2>r$ and
therefore is stable against small fluctuations. On the contrary
the second, phase mode, becomes unstable at $p>p_c=\sqrt{r/3}$. We
would like to stress that the relativistic nature of excitations
is crucial to get this instability. Despite being continuum
equations (\ref{gpr}) rely on the presence of the underlying
lattice, which breaks the translational invariance. Otherwise the
equations of motion would be Gallilean invariant and no critical
current would exist.

From the analysis above we see that close to the
superfluid-insulator phase transition current states become
unstable at momenta inversely proportional to the correlation
length of the condensate. As one goes deeper into the superfluid
regime the correlation length decreases saturating at one (in the
lattice units) and we come back to the Gross-Pitaveskii result of
instability occurring at $p=\pi/2\sim 1$.

\subsubsection{Incommensurate case}

It is also interesting to consider the stability of the current
states at the noncommensurate filling. In this case the system
remains superfluid at arbitrarily strong
interactions~\cite{sachdev_book}. If the interactions are weak the
system is in the Gross-Pitaveskii regime and the filling is not
important. In this case we expect a usual modulational instability
at $p\approx \pi/2$. At the same time, when the interaction
strength becomes very large, we can think about excess particles
as hard-core bosons moving on top of the Mott insulator. But in
turn, this would be equivalent to a spin one half system. At the
mean field level we can use again spin-wave theory to see that
$p_c$ is exactly equal to $\pi/2$. This suggests that $p_c$ should
have a minimum at the intermediate interaction strength.

The solution of Eq.~(\ref{gpr}) corresponding to the
noncommensurate filling factor can be written as:
\be
\psi(x,{\bf z},t)=\rho\,\mathrm e^{i p x+i\mu t},
\ee
where $\rho=\sqrt{r+\mu^2-p^2}$ and $\mu$ is related to the
deviation from integer filling $\delta n$:
\be
\delta n=A \mu\rho^2.
\label{delta_n}
\ee
As in the commensurate case, in the long wavelength
limit there are two branches describing a gapped amplitude mode
and gapless phase fluctuations. The dispersion of the latter for
${\bf q}$ parallel to the x-axis reads
\be
\omega(q)={2\mu p\over 2\mu^2+\rho^2}\, q+{\rho \over
2\mu^2+\rho^2}\sqrt{3\rho^2-2r}\,|q|.
\ee
From this we observe that the current state first becomes unstable
when $3\rho^2=2r$. Together with (\ref{delta_n}) this gives the
critical momentum
\be
p_c=\sqrt{\frac{r}{3}+{9\over 4r^2}{\delta n^2\over A^2}}.
\label{pc}
\ee
This result reduces to the commensurate limit for $\delta n=0$.
However, for any nonzero $\delta n$ it shows that $p_c$ reaches
the minimum value $p_c^\star\propto \delta n^{1/3}$ at $r_0\propto
\delta n^{2/3}$ and then diverges as $r\to0$. While the
divergence is the spurious result of the continuum approximation;
it should be cut of by the lattice at $p\approx 1$, the existence
of the minimum agrees with the simple general argument given
above.

\subsection{Gutzwiller approximation}

Having derived the conditions for the stability of the
current-carrying condensate in a lattice in the two extreme
limits, one can try to find a unifying approach, which
interpolates between them. A natural choice is the Gutzwiller
approximation. This is a time-dependent generalization of the
standard mean field theory, where the wavefunction is assumed to
be factorizable:
\be
\ket{G}=\prod_j \left[\sum_{n=0}^\infty f_{jn}\ket{n}_j\right].
\label{G}
\ee
Here $j$ denotes a site index and $n$ site occupation. The ansatz
(\ref{G}) supplemented by self-consistency conditions leads to
equations of motion:
\begin{eqnarray}
&&-i\dot{f}_{jn}=\frac{U}{2} n(n-1)f_{jn}-\nonumber\\
&&-Jz(\sqrt{n} f_{j, n-1}\psi_j +\sqrt{n+1}
f_{j,n+1}\psi^\star_j),
\label{TDG}
\end{eqnarray}
where
\be
\psi_j\equiv{1\over z}\sum_{i\in O}\bra{G}a_i\ket{G},
\label{psij}
\ee
$O$ is the set of nearest neighbors to $j$ and $z$ is the
coordination number ($z=2d$ for a hypercubic lattice). In practice
the sum in (\ref{G}) is limited to a finite number of states, on
each site, so that only a finite number of equations need to be
solved. We checked that in our numerical simulations we take
sufficient number of states so that the results are insensitive to
the truncation. In particular, for $N=1$ we compared simulations
for the spectrum truncated at $5$ and $10$ states per site and the
results were practically indistinguishable. The Gutzwiller
approximation can be justified at high dimensions, where the
coordination number $z$ becomes large. In this sense it is
reminiscent to the dynamical mean-field theory~\cite{kotliar}. In
reality, it is necessary to calculate first quantum corrections to
the evolution governed by Eq.~(\ref{TDG}) to see the validity of
the Gutzwiller result at a given dimensionality. We will postpone
such an analysis until the next section.

\begin{figure}[ht]
\includegraphics[width=8cm]{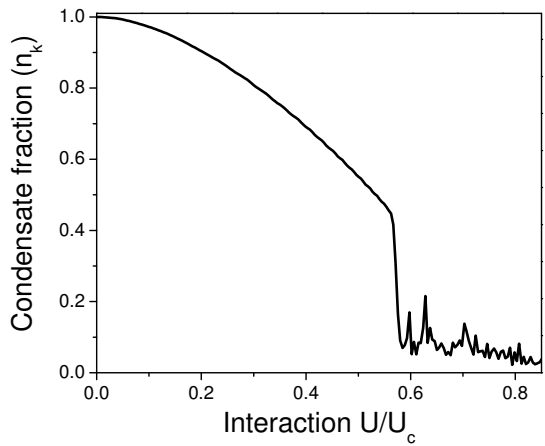}
\caption{Condensate fraction as a function of scaled interaction
for a two-dimensional condensate with filling $N=1$ evaluated
within Gutzwiller approximation on a square lattice of size
$80\times 2$~[\onlinecite{footote}]. The initial momentum is
$p=\pi/5$. The other parameters are $J=1$ (so that the interaction
strength corresponding to the transition to the Mott state is
$U_c\approx 23.2$) and the interaction increases in time as
$U=0.04 t$. The condensate motion clearly becomes unstable at
certain interaction ($U/U_c\approx 0.57$), which marks the
dynamical transition.}
\label{fig_inst}
\end{figure}

To find numerically the position of the dynamical instability
corresponding to Eq.~(\ref{TDG}) we can carry out one of the following
procedures. (i) Starting from the noninteracting state ($U=0$),
where the Gutzwiller ansatz becomes exact, and a given phase
gradient $p$ we adiabatically increase $U$. Observing either the
current or the condensate fraction (which we define as the
population of the state with the momentum $p$) we can identify the
critical interaction at which the motion becomes unstable (see
Fig.~\ref{fig_inst}).
(ii) Alternatively we can find numerically the mean field ground
state corresponding to given $U$ and $J$ and adiabatically
increase a gauge potential so that (\ref{psij}) modifies to
\beq
&&\psi_{j,{\bf l}}\equiv{1\over z} \biggl(\bra{G}a_{j+1,{\bf
l}}\ket{G}\mathrm e^{i\phi(t)}\nonumber\\
&&+\bra{G} a_{j-1,{\bf l}}\ket{G}\mathrm e^{-i\phi(t)}+\sum_{{\bf
l}^\prime\in O^\prime}\bra{G}a_{j,{\bf
l^\prime}}\ket{G}\biggr).\phantom{XX} \label{psij1}
\eeq
Here we explicitly introduced indices along the current $j$ and in
the transverse direction ${\bf l}$; $O^\prime$ is a subset of $O$,
which excludes the sites $\{j\pm 1,{\bf l}\}$. If the system is
stable then the condensate remains at the momentum $p=0$ in a
moving lattice, which is of course equivalent to a moving
condensate with $p=\phi(t)$ in a stationary lattice. Once the
motion becomes unstable the distribution at $p=0$ rapidly drops.
The second approach is favorable, because it does not require
quantization of the momentum in units of $2\pi/M$ in finite
systems of longitudinal size $M$, where the actual calculations
are performed.

\begin{figure}[ht]
\includegraphics[width=8cm]{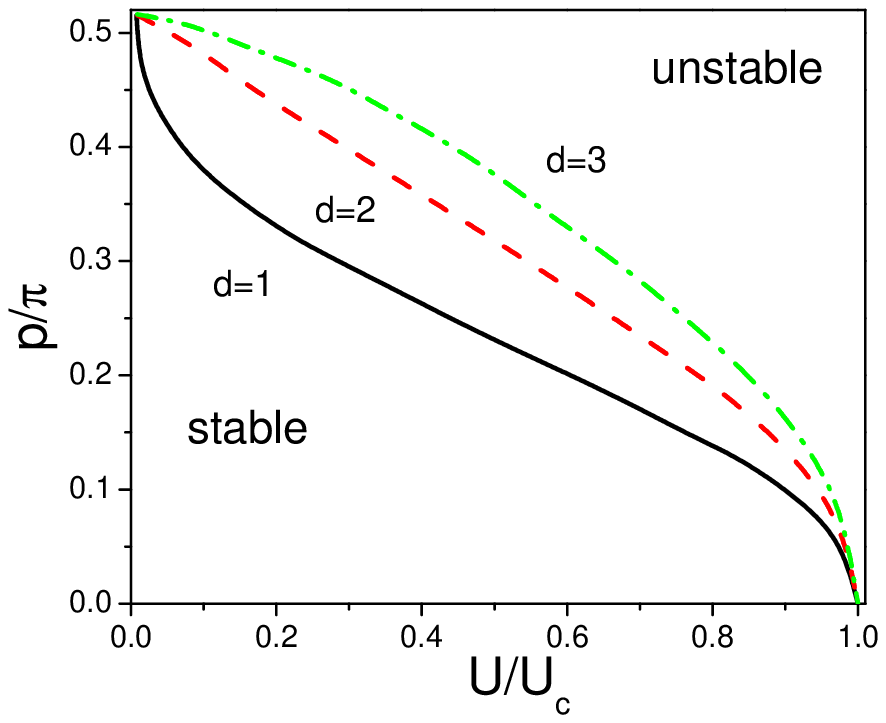}
\includegraphics[width=8cm]{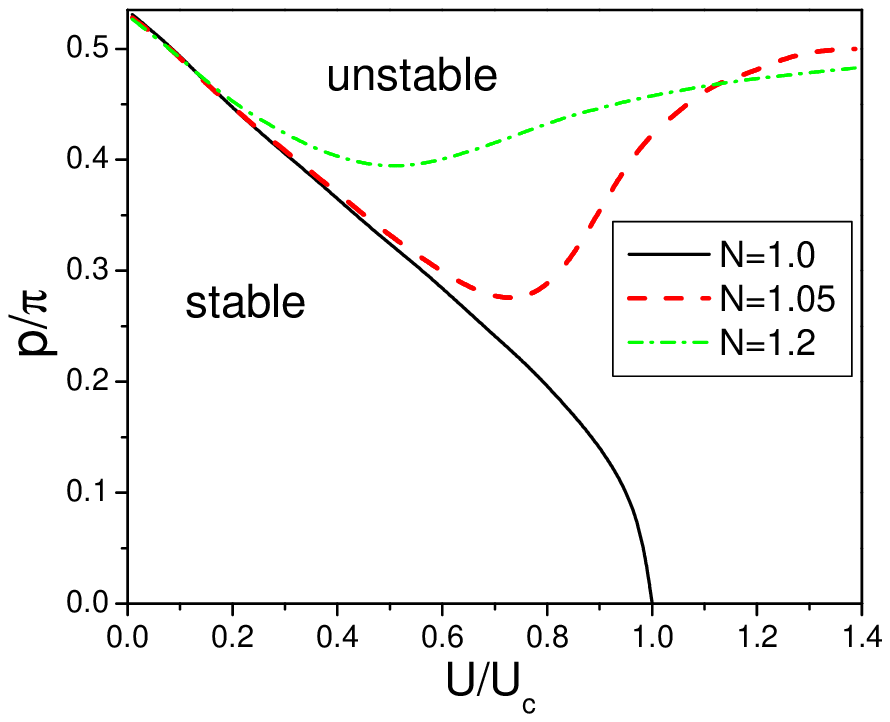}
\caption{Mean field phase diagram separating stable and unstable
motion of condensate. The vertical axis is the condensate momentum
in the inverse lattice units and the horizontal axis is the
normalized interaction. The top graph shows the result for integer
filling $N=1$ at different spatial dimensions. The bottom graph
describes a two-dimensional lattice with different filling
factors.}
\label{ph-dg2}
\end{figure}

Identifying the point of dynamical instability as described above
for different interaction strengths we can construct a mean field
phase diagram separating stable and unstable condensate motion for
both integer and noninteger filling factors (see
Fig.~\ref{ph-dg2}). This phase diagram is in complete agreement
with what we expected from the analysis given in the previous
subsections. Thus at small interactions the critical momentum
approaches $\pi/2$ for any filling or dimensionality of the
lattice. At integer filling the critical momentum vanishes at the
point of commensurate superfluid-insulator transition. In the
incommensurate case the critical momentum first goes down with the
interaction strength and then increases back to $\pi/2$ in the
strongly interacting regime.

\section{Beyond mean field theory. Current decay due to fluctuations.}
\label{sec2}

The analysis given in the previous section is valid only at the
mean-field level. Quantum and thermal fluctuations can destroy the
current by either phase slip tunneling or thermal activation. The
main goal of this section is to derive leading contributions to
the decay rate and check the validity of the mean field results.
To simplify the analysis we will concentrate on the two tractable
limits: the Gross-Pitaveskii regime describing the system deep in
the superfluid phase and the Ginzburg-Landau regime, which is
valid in the vicinity of the superfluid insulator transition,
where the correlation length becomes large compared to the lattice
constant. Also for simplicity we assume that the filling is large
and integer so that one can use the quantum-rotor model.

\subsection{Gross-Pitaveskii regime.}

\subsubsection{Current decay due to quantum tunneling}

As we argued above, the current state with $p<\pi/2$ is stable
with respect to small fluctuations. However, this state does not
correspond to the energy minimum, which has no current. So we
conclude that such a state must be metastable. Contrary to a
uniform system, where the momentum can be gauged away via the
Gallilean transformation, in a lattice there is a preferred
reference frame. This immediately implies that such a metastable
state should be able to spontaneously decay because of quantum
tunneling.

In the leading order in $U/JN$, which plays the role of the
effective Planck's constant for this problem~\cite{ap1}, the
tunneling rate ($\Gamma$) corresponds to the action ($S$) of the
bounce solution (instanton) of the classical equations of motion
in the inverted potential~\cite{coleman}:
\be
\Gamma\propto \mathrm e^{-S}.
\ee
We will not attempt to compute the prefactor here and will
concentrate only on $S$. We just point out that the prefactor
scales with the system size and in the thermodynamic limit the
tunneling rate per unit volume is size independent. Explicitly the
action reads as:
\beq
&&S=\sum_{j,{\bf l}} \int d\tau \biggr[ {1\over
2U}\left({d\phi_{j,{\bf l}}\over d\tau}\right)^2
-2JN\cos(\phi_{j,{\bf l+1}}-\phi_{j,{\bf l}})\nonumber\\
&&~~~~-2JN\cos(p+\phi_{j+1,{\bf l}}-\phi_{j,{\bf l}})\biggr],
\label{s8}
\eeq
where as before $j$ denotes the coordinate along the current
direction and ${\bf l}$ corresponds to the transverse coordinates.
In (\ref{s8}) we redefined phases compared to (\ref{gp2}):
$\phi_j\to \phi_j-px_j$ so that $\phi_j=0$ corresponds to a
metastable current state and the new phases are the fluctuations
around this state. It is convenient to redefine the imaginary time
$\tau$, measuring it in Josephson time units: $\tau\to
\tau/\sqrt{UNJ}.$ Then it is easy to see that
\be
S=\sqrt{JN\over U}s,
\label{s5}
\ee
where
\beq
&&s=\sum_{j,{\bf l}} \int d\tau \biggl[{1\over
2}\left({d\phi_{j,{\bf l}}\over
d\tau}\right)^2-2\cos(\phi_{j,{\bf l+1}}-\phi_{j,{\bf l}})\nonumber\\
&&~~~~~~~~~~~~~~~~~~~~~-2\cos(p+\phi_{j+1,{\bf l}}-\phi_{j,{\bf
l}})\biggr].
\label{s2}
\eeq
The desired instanton trajectory is the solution of the
Euler-Lagrange equations, which can be obtained extremizing the
action with respect to the phases $\phi_{j,{\bf l}}$ and subject
to a boundary conditions: $\phi_{j,{\bf l}}=0$ at $\tau=\pm\infty$
and at $|{\bf l}|,|j|=\infty$.

Before proceeding with a general analysis in higher spatial
dimensions let us consider the case $d=1$ first. In this work we
are interested in the decay close to the critical current
$p_c=\pi/2$. Clearly as $p\to p_c$ the effective tunneling barrier
becomes weaker and weaker and hence we can expand (\ref{s2}) in
powers of $\phi_j$:
\beq
&&s\approx \sum_{j} \int d\tau \biggl[{1\over
2}\left({d\phi_{j}\over
d\tau}\right)^2\nonumber\\
&&~~~~~~~~+\cos( p)\,
(\phi_{j+1}-\phi_{j})^2-{(\phi_{j+1}\!-\phi_{j})^3\over 3}\biggr].
\label{s6}
\eeq
This expansion is similar to that used in the analysis of
thermally mediated decomposition near spinodal point~\cite{klein}.
In the action above we used that $\sin p\approx \sin p_c=1$. Now
we can do another rescaling:
\be
\phi_j=\cos(p)\,\tilde \phi_j, ~~\tau={\tilde\tau\over
\sqrt{\cos(p)}},
\label{scaling}
\ee
which simplifies the action even further:
\be
s\approx (\cos p)^{5/2}\tilde s,
\label{s3}
\ee
where
\be
\tilde s=\sum_{j} \int d\tilde\tau {1\over
2}\left({d\tilde\phi_{j}\over
d\tilde\tau}\right)^2\!\!\!+(\tilde\phi_{j+1}-\tilde\phi_{j})^2\!-{
(\tilde\phi_{j+1}-\tilde\phi_{j})^3\over 3}.
\label{ss}
\ee
Note that $\tilde s$ is just a number of the order of one so that
(\ref{s3}) completely determines the momentum dependence of the
action. The scaling (\ref{scaling}) guarantees that the original
phases $\phi_j$ remain small for the instanton trajectory and
(\ref{s6}) is indeed a correct asymptotical form of (\ref{s2}).

Since the action $\tilde s$ contains no small parameters, the
bounce solution should be localized within a few sites. Without
loss of generality we can assume that the maximum phase difference
develops between the sites labeled by $j=0$ and $j=1$. For the
rest of the system the phase gradients are relatively small so we
can neglect cubic terms. Then those degrees of freedom can be
integrated out:
\be
\tilde\phi_j(\tau)=\pm \int {d\omega\over
4\pi}\alpha(\omega)\mathrm e^{-i\omega\tau}\lambda(\omega)^{j-1},
\label{s9}
\ee
where $\alpha(\tau)=\tilde\phi_1(\tau)-\tilde \phi_0(\tau)$ and
$\alpha(\omega)$ is its Fourier transform;
\be
\lambda(\omega)=1+{\omega^2\over 4}-{\omega\over
\sqrt{2}}\sqrt{1+{\omega^2\over 8}}.
\ee
Substituting (\ref{s9}) into (\ref{ss}) we find:
\beq
&&\tilde s\approx \int d\tau\left[{1\over
4}\left({d\alpha(\tau)\over d\tau}\right)^2+\alpha^2(\tau)-{1\over
3}\,\alpha^3(\tau)\right]\nonumber\\
&&~+\int {d\omega \over 4\pi} |\alpha(\omega)|^2\, {|\omega|\over
|\omega|+\sqrt{8+\omega^2}}.
\label{s1}
\eeq
Clearly, if we ignore the last term in the expression (\ref{s1})
we get the action of a single particle moving in a metastable
potential. The last term represents a dissipative-like
contribution coming from the rest of the chain. If we ignore this
term then the solution extremizing (\ref{s1}) is:
\be
\alpha(\tau)={3\over \cosh^2\,{t/\sqrt{2}}}.
\ee
This yields $\tilde s=24/5\equiv 4.8.$ In a general case with
dissipation we can try a variational ansatz solution:
\be
\tilde\alpha(\tau)={A\over \cosh^2\,r\tau}.
\ee
A simple numerical analysis gives
\be
r\approx 0.64,\; A\approx 3.31,\; \tilde s\approx 7.11.
\label{s7}
\ee
So the action is about a factor of $1.5$ larger than without the
bath degrees of freedom. Using (\ref{s9}) and (\ref{s7}) one can
verify the consistency of the harmonic approximation used for the
sites other than "1" and "0". In particular, it is straightforward
to get:
\be
{|\tilde\phi_2(0)-\tilde\phi_1(0)|\over \alpha(0)}\approx 0.326,
\ee
which is relatively small. The difference between phases in
further nearest neighbor sites is even less. Therefore for them
the harmonic approximation is justified even better.

Instead of harmonic treatment of the phases other than
$\tilde\phi_1$ and $\tilde \phi_2$ we can exactly take into
account the four nearest neighbor sites and ignore the rest of the
chain. Then the instanton solution is parametrized by the two
independent angles $\alpha$ and $\beta$:
\be
\tilde\phi_0=-\tilde\phi_{-1}=\alpha/2,\;
\tilde\phi_{1}=-\tilde\phi_{-2}=\alpha/2+\beta.
\ee
Substituting this into the action and solving the corresponding
Euler-Lagrange equations one can show that in this case $\tilde
s\approx 6.1$, which is about a factor of $1.26$ larger than the
result with $\beta\equiv 0$. This number is the exact lower bound
for the action $\tilde s$ since the other degrees of freedom can
only increase the action. We will not further attempt to improve
the accuracy of $\tilde s$ noting only that the variational result
$\tilde s\approx 7.1$ should be very close to the exact value.

We can straightforwardly generalize the one-dimensional results to
higher spatial dimensions. In particular, using the same arguments
as before close to the critical current we can expand the action
(\ref{s2}) up to the cubic order in $\phi_{j,{\bf l}}$:
\beq
&&s\approx \sum_{j,{\bf l}} \int d\tau \biggl[{1\over
2}\left({d\phi_{j_{\bf l}}\over
d\tau}\right)^2+\cos(p)\,(\phi_{j+1,{\bf l}}-\phi_{j,{\bf
l}})^2\nonumber\\
&&~~~~~~+\sum_{\bf l^\prime\in O^\prime}(\phi_{j,{\bf
l^\prime}}-\phi_{j,{\bf l}})^2-{1\over 3}(\phi_{j+1,{\bf
l}}-\phi_{j,\bf l})^3\biggr].
\label{s6a}
\eeq
Note that at $p\to\pi/2$ only longitudinal modes become soft,
acquiring a prefactor $\cos p$ in front of the quadratic term in
the action. This implies that the transverse width of the
instanton should be much larger than its longitudinal size and we
can safely use the continuum approximation for the phases along
the transverse directions. Then instead of (\ref{s6a}) we derive:
\beq
&&s\approx \sum_{j} \int d\tau d^{d-1} x \biggl[{1\over
2}\left({d\phi_{j}\over d\tau}\right)^2 +\left({d\phi_j\over d{\bf
x}}\right)^2\nonumber\\
&&~~~+\cos(p)\,(\phi_{j+1} -\phi_{j})^2-{1\over
3}(\phi_{j+1}-\phi_{j})^3\biggr].
\label{s6b}
\eeq
In this equation ${\bf x}$ denotes transverse coordinates which
reside in a $d-1$ dimensional space. As in the one dimensional
case we can rescale
\be
\phi=\cos (p)\, \tilde \phi,\; \tau={\tilde \tau\over
\sqrt{\cos(p)}},\; x={\tilde x \sqrt{2}\over\sqrt{\cos p}}.
\label{scaling1}
\ee
In this way the action (\ref{s6b}) becomes:
\be
s\approx (p_c-p)^{6-d\over 2}\tilde s_d,
\label{s6c}
\ee
where
\beq
&&\tilde s_d=2^{d-1\over 2}\sum_j\int d^d\zeta \, \Biggl[{1\over
2}\left({d\tilde\phi_j\over d{\bf \zeta}}\right)^2\nonumber\\
&&~~~~~~~~~~~+ (\tilde\phi_j-\tilde\phi_{j+1})^2 -{1\over
3}(\tilde\phi_j-\tilde\phi_{j+1})^3\Biggr].
\eeq
Here ${\bf\zeta}=(\tilde\tau,\tilde{\bf x})$ is a $d$-dimensional
space-time coordinate. As before $\tilde s_d$ is just a number,
which depends only on dimensionality. The result (\ref{s6c}) is
quite remarkable. First of all it shows that the action in higher
dimensions vanishes much more slowly near the critical current. From
the scaling (\ref{scaling1}) it is obvious that the characteristic
transverse dimension of the instanton scales as $1/\sqrt{p_c-p}\gg
1$, justifying the continuum approximation. Above $d=6$ the
tunneling action would experience a discontinuous jump at $p=p_c$,
however since we deal with $d\leq 3$ this is not relevant. Now let
us try to estimate $\tilde s_d$. We again proceed in the same
spirit as in the one-dimensional case. In the first approximation
we consider only a single phase slip $\tilde\phi_1=\alpha/2$,
$\tilde\phi_2=-\alpha/2$ and treat the motion of other phases in
the harmonic approximation. The corresponding dimensionless action
reads
\beq
&&\tilde s_{d}=2^{d-1\over 2}\int d^d\zeta\left[ {1\over
4}(\nabla\alpha)^2+\alpha^2-{1\over 3}\alpha^3\right]\nonumber\\
&&+{1\over 2}\int {d^d k \over (2\pi)^d} |\alpha({\bf k})|^2\,
{k\over k+\sqrt{8+k^2}},
\label{sd}
\eeq
where $\alpha({\bf k})$ is the Fourier image of $\alpha({\bf x})$.
If we ignore the last dissipative term in (\ref{sd}), then our
action is identical to that considered in
Ref.~[\onlinecite{coleman}] for the decay of a false vacuum. In
that work it was argued that the bounce solution is spherically
symmetric and it satisfies the following equations of motion:
\be
{1\over 2}{1\over\zeta^d}{d\over d\zeta}\left(\zeta^d{d\alpha\over
d\zeta}\right)=2\alpha-\alpha^2,
\label{traj}
\ee
with the boundary conditions: $\alpha(\infty)=0$ and
$\alpha^\prime(0)=0$. These equations can be easily solved
numerically and the result is:
\be
\tilde s_2\approx 21.92,\; \tilde s_3\approx 87.32.
\ee
So it is clear that at higher dimensions not only the exponent of
$(p_c-p)$ in the instanton action gets lower but also the
numerical prefactor gets larger. Now we can find the correction to
the action due to the bath degrees of freedom coming from the last
term in (\ref{sd}). Instead of using the variational approach as
we did in the one dimensional case, we will use the exact solution
of (\ref{traj}) to evaluate the contribution of the bath term in
the action. This will be the {\em exact} upper bound of the action
$\tilde s_d$. Direct evaluation of (\ref{sd}) gives
\be
4.8< \tilde s_1< 8.0,\; 21.9< \tilde s_2< 27.6,\; 87.3< \tilde s_3
< 97.5.
\ee
Obviously the local approximation $\phi_j=0$ for $j\neq 0,1$ works
better and better at higher dimensions implying that the effective
size of the instanton in the longitudinal direction (along the
current) decreases with the dimensionality of the space.

\begin{figure}[ht]
\includegraphics[width=8.5cm]{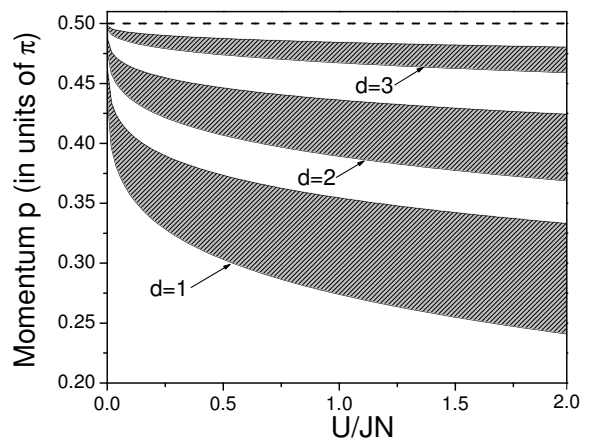}
\caption{Large $N$ zero-temperature phase diagram for the
nonequilibrium superfluid-insulator transition. The dashed line
represents mean field transition at $p=\pi/2$. The shaded regions
correspond to the tunneling action satisfying $1\leq S\leq 3$,
obtained within the discrete phase model in different spatial
dimensions. Below the shaded regions the tunneling action is large
and the current decay is slow, so the superfluid state is stable
for long time scales. Above the shaded regions the current decays
very fast and the superfluid state is unstable. }
\label{ph-dg}
\end{figure}

Because the decay rate is strongly (exponentially) dependent on
momentum and coupling constants we can approximately define the
stable phase at which the tunneling action is larger than some number,
say $S>3$ and unstable phase, when there is no exponential
suppression of the tunneling of phase-slips, say $S<1$. The region
in between will denote the crossover between the stable and
unstable regimes. In this way we can define a crossover phase
diagram (Fig.~\ref{ph-dg}).
We see that except for $U/JN\ll 1$, there is a very strong
broadening of the classical transition in one dimension. On the
contrary in the three-dimensional case effects of quantum
fluctuations are relatively weak and the crossover is very sharp.
We would like to point out that the derivation given here is valid
only deep in the superfluid regime $U/JN\ll 8d$. Close to the
critical point it is necessary to use the coarse-grained
description which we discuss later.

To summarize this subsection we write explicit expressions for the
tunneling action in the phase model in three spatial dimensions:
\beq
&&S_{1d}\approx 7.1\sqrt{JN\over U}\left({\pi\over
2}-p\right)^{5/2},\label{s_1d}\\
&&S_{2d}\approx 25\sqrt{JN\over U}\left({\pi\over
2}-p\right)^2,\label{s_2d}\\
&&S_{3d}\approx 93\sqrt{JN\over U}\left({\pi\over
2}-p\right)^{3/2}\label{s_3d}.
\eeq

\subsubsection{Thermally activated current decay.}

Now let us turn our attention to broadening of the mean field
transition due to thermal fluctuations. A general formalism for
finding the decay rate was developed by Langer~\cite{langer}. It
was later successfully applied to quasi-one-dimensional
superconductors~\cite{langer1,halperin} and to three dimensional
superfluids at small currents~\cite{lf}.

Before proceeding with this general method let us point out an
essential difference between conventional condensed matter systems
and the cold atoms systems addressed in this work. In the former,
it is the environment, which introduces thermal noise and
dissipation~\cite{caldeira} to the system leading to diffusion in
energy space and eventually thermal activation. In cold atom
systems by contrast, the temperature is introduced into the system
during the preparation of the condensate. I.e., the initial state
of the condensate is described by a thermal density matrix rather
than a pure wave function. Later the system is essentially
isolated from the environment and evolves according to the
Hamiltonian equations of motion. We point out that the formalism
of Ref.~[\onlinecite{langer}] was based on very general
assumptions, and therefore should be independent of the details of
the thermal fluctuations. Nevertheless, certain issues arise that
require special attention. Consider for example a superfluid
flowing in a container whose walls act as the thermal bath. The
wall as well as the thermal fluctuations arising from it set a
preferred reference frame, breaking the galilean invariance of the
superfluid and thus allowing for the current decay. An isolated
superfluid, on the other hand, even if prepared at finite
temperature, is galilean invariant. Thus current in such a
superfluid cannot decay unless there is an external potential,
such as a lattice, that sets a preferred reference frame. Because
we are interested in the current decay in the limit where the
lattice is strong and BHM is applicable, this subtlety is
irrelevant for our consequent analysis. The effect of breaking of
galilean invariance by weak external potential on thermally
activated current decay in one dimensional superconductors was
recently studied in~Ref.~[\onlinecite{Khlebnikov}].

To simplify the analysis we assume that the temperature is small
so that there is no difference between the energy and the free
energy. Indeed, the difference between the two amounts to the
product $T\mathcal S$, where $\mathcal S$ is the entropy of the
system. At small temperatures the latter vanishes in the
superfluid as $T^d$ so that the product $T\mathcal S$ goes to zero
at least as $T^2$ at $T\to 0$ and is always negligible compared to
the activation energy, which does not depend on temperature.

As in the previous subsection, we first consider the one
dimensional case. Following Refs.~[\onlinecite{langer, halperin}]
we find the stationary solutions of the classical equations of
motion:
\be
{1\over 2}{d^2\phi_j\over
d\tau^2}=\sin(p+\phi_{j+1}-\phi_j)+\sin(-p+\phi_{j-1}-\phi_j).
\label{eq7}
\ee

\begin{figure}[ht]
\includegraphics[width=8cm]{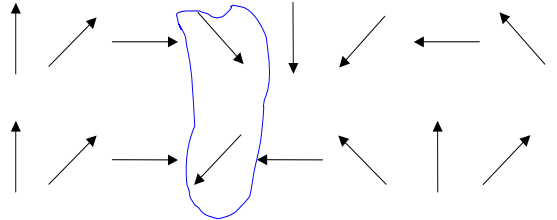}
\caption{Schematic representation of a metastable current carrying
state and an unstable saddle point solution. The arrows represent
the superfluid phase at different sites of the lattice.}
\label{fig1}
\end{figure}

Clearly $\phi_{j}= 0$ describes a metastable state carrying the
current $2JN\sin p$. We note again that phases in (\ref{eq7}) are
the deviations from the metastable current state. The other saddle
point solution separating the states with different currents is
\be
\phi_j=\left\{\begin{array}{ll} (p_\pm^\prime-p) j & j<0\\
\pi+p_\pm^\prime (j-2)-pj & j\geq 1\end{array}\right.,
\label{eq8}
\ee
where $p^\prime_-\approx p-(\pi-2 p)/M$ and $p^\prime_+\approx
p+(\pi+2 p)/M$ for a periodic chain of size $M$. The indices
``$-$'' and ``$+$'' correspond to phase slip and anti-phase slip,
respectively, with the convention that the slip reduces the
current. Schematically the saddle point and the metastable
solutions are depicted in Fig.~\ref{fig1}.
Clearly the energy difference between the two states is
\be
\Delta E_-=2JN(2\cos p\,-\sin p\,(\pi-2p))
\ee
and
\be
\Delta E_+=2JN(2\cos p\,+\sin p\,(\pi+2p)).
\ee
Correspondingly the decay rate to lower (higher) current state is
proportional to
\be
\Gamma_{\mp}\propto \mathrm e^{-\beta\Delta E_{\mp}}=\mathrm
e^{-2JN\beta(2\cos p\mp (\pi\mp 2p)\sin p)}.
\label{s12}
\ee
In particular when $p\to \pi/2$ we have:
\be
\Gamma_{-}\propto\mathrm e^{-{4\over 3}NJ\beta (\pi/2-p)^3}.
\label{g-}
\ee
In the one dimensional case it is also straightforward to evaluate
the prefactor in the decay rate. We give details of such
derivation in the Appendix~\ref{seca1} and quote only the final
result here:
\begin{widetext}
\be
\Gamma_-\approx {64 \cos p\over \pi\tau}\sqrt{JN\over U}
\exp\left(-{\pi/2-p\over 4}\tan p-{4NJ\over T}(\cos
p\,-(\pi/2-p)\sin p)\right).
\label{gam1}
\ee
\be
\Gamma_+\approx {64 \cos p\over \pi\tau}\sqrt{JN\over
U}\exp\left({\pi/2+p\over 4}\tan p-{4NJ\over T}(\cos
p\,+(\pi/2+p)\sin p)\right).
\label{gam2}
\ee
\end{widetext}
Here $\tau$ is a relaxation time of the condensate towards thermal
equilibrium, which we will leave as a phenomenological parameter.

We can now compare the leading exponential terms for the two
current decay mechanisms. Thus if we require that the exponent in
(\ref{s12}) is equal to the tunneling action computed in the
previous section we can find when the two exponents coincide. It
is convenient to introduce a characteristic temperature scale
$T_Q$, at which the energy of the zero-point fluctuations is equal
to the thermal energy of the corresponding classical system. Using
Bogoliubov's approximation one finds that
\be
T_Q={2\sqrt{2}\over \pi}\sqrt{NJU}.
\label{tq}
\ee
We can now rewrite the expression for $\Gamma_-$ at $p$ close to
$p_c=\pi/2$ in a more convenient form:
\be
\Gamma_-\approx {64\mathrm  \cos p\over
e^{1/4}\pi\tau}\sqrt{JN\over U} \exp\left[-{\sqrt{2}\pi\over
3}{\sqrt{NJ\over U}}{T_Q\over T}(p_c-p)^{3}\right].
\ee
Note that the exponent in the expression above coincides with that
in (\ref{s_1d}) when
\be
T^\star\approx 0.21\,T_Q\sqrt{{\pi/ 2}-p}.
\label{1d}
\ee
The temperature $T^\star$ separates regimes of thermally activated
and quantum current decay. Thus if $T<T^\star$ thermal phase slips
are unimportant so that quantum tunneling dominates the decay and
vice versa. Note that unless $p$ is very close to $\pi/2$, the
crossover temperature $T^\star$ is of the order of the
characteristic Josephson energy $T_Q$ (or equivalently the sound
velocity in the lattice units). Under present experimental
conditions it is very easy to achieve $T\ll T_Q$ and thus
$T<T^\star$ and therefore to observe current damping due to
quantum phase slips.

In higher dimensions we can not find an explicit analytic
expression for the energy of the metastable state. However, we can
get an approximate result near the critical current. Using again
the idea that the transverse fluctuations can be treated in the
continuum approximation and expanding $\cos(p+\phi_{j+1}-\phi_j)$
up to the third order in phases we can write the energy in the
approximate form:
\beq
&&E_d\approx JN\sum_j \int d^{d-1} x \biggl[\left({d\phi_j\over
dx}\right)^2\nonumber\\
&&~~~~~~+\cos (p)\, (\phi_{j+1}-\phi_j)^2-{1\over
3}(\phi_{j+1}-\phi_j)^3\biggr],
\eeq
where $\phi_j(x)$ is the nontrivial solution of the corresponding
Euler-Lagrange equations vanishing at $x\to\infty$. After
rescaling $\phi=\cos(p)\,\tilde\phi$ and $x=\tilde x
\sqrt{2}/\sqrt{\cos p}$ we find:
\beq
&&E_d\approx JN\, 2^{d-1\over 2}\left(p_c-p\right)^{7-d\over
2}\sum_j\int d^{d-1}\tilde x
\nonumber\\
&&\biggl[ {1\over 2}\left({d\tilde\phi_j\over d\tilde
x}\right)^2\!\!\!+(\tilde\phi_{j+1}-\tilde\phi_j)^2-{1\over
3}(\tilde\phi_{j+1}-\tilde\phi_j)^3\biggr].
\eeq
Note that the integral in the expression above coincides with
$\tilde s_{d-1}$ up to a number $2^{d-2\over 2}$. So using results
(\ref{s_1d}), (\ref{s_2d}), and (\ref{g-}) we immediately conclude
that
\beq
&&E_1\approx 1.3\,JN \left({\pi\over 2}-p\right)^3, \\
&&E_2\approx 10\, JN  \left({\pi\over 2}-p\right)^{5\over
2},\\
&&E_3\approx 35 JN \left({\pi\over 2}-p\right)^2.
\eeq
Correspondingly the exponents in the thermal and quantum decay
rates become the same at a temperature
\be
T^\star\approx 0.44\, T_Q \sqrt{{\pi/ 2}-p}
\ee
both in two and three dimensions. This crossover temperature is
about a factor of two higher than  in the one dimensional case
(\ref{1d}). If $p$ is not too close to $\pi/2$, $T^\star$ is again
of the order of $T_Q$ and thus the quantum tunneling should be
responsible for the current damping below the mean field
transition at experimentally relevant temperatures.

\subsection{Current decay in the vicinity of the SF-MI phase
transition.}

As we show in the Appendix~\ref{sec:ax} the quantum action in
imaginary time takes the following form:
\be
S=C\int d^dz dx\; \left|{d\psi\over d{\bf
z}}\right|^2+\left|{d\psi\over dx}\right|^2-|\psi|^2+{1\over
2}|\psi|^4.
\label{gl}
\ee
Here $z$ denotes the imaginary time and transverse coordinates
which form a $d$-dimensional space, $x$ is a longitudinal
coordinate along the current. Note that in this section we measure
coordinates in units of the coherence length. This is because we
focus on the commensurate case and hence we are interested only in
the superfluid regime $r>0$ (see Eq.~(\ref{gpr})). In this case it
is convenient to rescale $x\to x/\sqrt{r}$ to simplify the
notations (in the original lattice units $x$ is measured in the
units of the correlation length $\xi$). The constant $C$ depends
on the original microscopic parameters of the underlying
Hamiltonian. Within the variational ansatz described in
Appendix~\ref{sec:ax} we find~\cite{ehud}:
\be
C={1\over u}{1\over  2(2d)^{d/2}}\left(1-u\right)^{3-d\over 2},
\label{c}
\ee
where $u=U/8dJN$ is the dimensionless interaction;
\be
\xi={1\over \sqrt{2d \left(1-u\right)}}.
\label{xi}
\ee

In the case when thermal fluctuations are more important than the
zero point motion, we are interested in the energy functional
rather than the action:
\be
\mathcal E=C^\prime\int d^{d-1}z dx\; \left|{d\psi\over d{\bf
z}}\right|^2+\left|{d\psi\over dx}\right|^2-|\psi|^2+{1\over
2}|\psi|^4,
\label{gle}
\ee
where ${\bf z}$ comprises now $d-1$ transverse coordinates only.
The value of the constant $C^\prime$ can be found within the mean
field approximation  (see Appendix~\ref{sec:ax}):
\be
C^\prime=JN{1\over u (2d)^{d/2-1}}\left(1-u\right)^{2-d/2}.
\ee

Before proceeding we would like to point out that the mean field
expressions for $\xi$ and $C$ are valid in the vicinity of the
quantum phase transition at large spatial dimensions. For example,
at $d=1$ the superfluid insulator transition belongs to the
Kosterlitz-Thouless universality class and it is characterized by
proliferation of vortices~\cite{doniah,fradkin}. In particular,
dissipation in two spatial dimensions in the vicinity of the
thermal superfluid-to-normal fluid transition, which is also of
the Kosterlitz-Thouless class was studied in
Ref.~[\onlinecite{hn}]. It was shown that the dissipation comes
from unbinding of existing vortices and it does not have an
exponential suppression. So while the mean field description near
the quantum critical point in one dimension is questionable, we
believe that it is justified in two and especially in three
spatial dimensions.

Decay of superconducting current in the GL model was studied by
several authors. In particular, the exponent characterizing the
decay rate in the one dimensional case was studied in
Ref.~[\onlinecite{langer1}] and the prefactor setting the time
scale was later found in Ref.~[\onlinecite{halperin}]. In three
dimensions at small currents the corresponding exponent was
derived by Langer and Fisher~\cite{lf}, where it was shown that
the decay rate vanishes as $\exp(-C/p)$ as $p\to 0$. However, here
we are interested in quite the opposite limit $p\to p_c$, where
the method used in that paper does not work.

Let us start our analysis from the quantum decay. We first
emphasize that quantum in this context means due to fluctuations
beyond the saddle point approximation. Contrary to the
Gross-Pitaveskii regime, where the classical description is well
controlled by the smallness of the ratio $U/JN$, there is no
obvious small parameter here. The validity of the mean field
description in this case can be checked {\em a-posteriori} by
explicit computation of quantum corrections. The other comment we
would like to make is that the parameters $C$ and $\xi$ entering
the Ginzburg-Landau action are generally renormalized and deviate
from the mean field results.

To compute the tunneling action we need to find a bounce-solution
of the classical equations of motion in imaginary time. Instead of
using complex fields $\psi_j$ we introduce two real fields $\eta$
and $\phi$ describing amplitude and phase fluctuations around the
metastable minimum:
\be
\psi(x,{\bf z})=\sqrt{1-k^2}(1+\eta(x,{\bf z}))\mathrm
e^{ikx+i\phi(x,{\bf z})}.
\ee
Here we intentionally use another notation $k$ for the condensate
momentum, because it is measured in the units of inverse
correlation length $\xi$. It is related to the usual momentum $p$
in inverse lattice units as $k=p\,\xi$. We can expect that close
to the critical current both $\eta$ and $\phi$ remain small and we
can expand the action up to the cubic terms in these fields to
find the correct asymptotical behavior of the instanton action at
$k\to k_c$:
\begin{widetext}
\be
s\approx {2\over 3}\int d{\bf z} dx\; (\partial_{\bf
z}\eta)^2+(\partial_{\bf z}\phi)^2+(\partial_x\eta)^2
+(\partial_x\phi)^2+2(1-k^2)\eta^2+4k\eta\,\partial_x\phi+2\eta
(\partial_{\bf z}\phi)^2+2\eta (\partial_x\phi)^2+{2\over
\sqrt{3}}\eta^2\,\partial_x\phi+{4\over 3}\eta^3.
\label{gl1}
\ee
\end{widetext}
It is easier to start the analysis of (\ref{gl1}) calculating the
corresponding energy in $d=1$. This problem was already solved in
Ref.~[\onlinecite{halperin}] for all values of $k$. So we will use
the asymptotic of their expression at $k\to k_c$ to compare with
our result and thus check the validity of our scheme. It is easy
to see that upon the transformation:
\be
\eta\to \eta-{k\over 1-k^2}\partial_x\phi
\label{trans}
\ee
the cross term in $\eta$ and $\phi$ vanishes in the quadratic part
of (\ref{gl1}). Since the amplitude $\eta$ mode remains gapped at
$k=k_c$ it can be disregarded and the approximate energy (which is
equivalent to the action (\ref{gl1}) at $d=0$) takes the form:
\be
\epsilon\approx \int dx\, {1\over 2}(\partial^2_x\phi)^2+{2
\sqrt{3}}(k_c-k)(\partial_x\phi)^2-{2\over \sqrt{3}}(\partial_x
\phi)^3.
\label{e}
\ee
Upon rescaling:
\be
x\to {x\over 3^{1\over 4}\, 2\sqrt{k_c-k}},\; \phi\to \phi \,
{3^{3\over 4}\over 2}\sqrt{k_c-k}
\ee
we obtain
\be
\epsilon \approx {9\, 3^{1\over 4}}(k_c-k)^{5/2}\int
dx\,(\partial^2_x\phi)^2+(\partial_x\phi)^2-(\partial_x \phi)^3.
\label{en}
\ee
The last integral is just a number equal to the energy of the
saddle point, which is a nontrivial solution of the Euler-Lagrange
equations:
\be
-2{d^2 \phi^\prime\over dx^2}+2\phi^\prime-3\phi^{\prime\,2}=0,
\ee
where $\phi^\prime\equiv d\phi/dx$. This equation has a simple
solution:
\be
\phi^\prime(x)={1\over \cosh^2 x/2},
\ee
which after substitution into (\ref{en}) gives:
\be
\epsilon\approx {48\over 5}3^{1\over 4}(k_c-k)^{5/2}.
\ee
This result coincides with that derived in
Ref.~[\onlinecite{halperin}] at $k\to k_c$. Now we can proceed
with a general d-dimensional case. As in the previous section the
instanton action in $d+1$ dimensions is equal to the barrier
energy in $d$-dimensions. After performing the transformations
(\ref{trans}) and ignoring the gapped amplitude mode $\eta$ we
find:
\beq
&&s\approx \int d^dzdx\, {1\over 2}(\partial^2_{{\bf
z},x}\phi)^2+{2\over 3}(\partial_{\bf z}\phi)^2+{1\over
2}(\partial^2_x\phi)^2\nonumber\\
&&~~~~~~~~~~~~+{2 \sqrt{3}}(k_c-k)(\partial_x\phi)^2-{2\over
\sqrt{3}}(\partial_x \phi)^3.
\label{act0}
\eeq
It is easy to see that the rescaling required to make the action
independent of the momentum is:
\be
x\to {x\over 3^{1\over 4}\, 2\sqrt{k_c-k}},\; \phi\to \phi \,
{3^{3\over 4}\over 2}\sqrt{k_c-k},\; z\to {z\over 6(k_c-k)}.
\label{rescale}
\ee
Upon these transformations the first term in the action becomes
irrelevant and in the leading order in $k_c-k$ the action reads:
\be
s_d={3^{9-4d\over 4}\over 2^d}(k_c-k)^{5-2d\over 2}\int d{\bf z}dx\,
(\nabla\phi)^2+(\partial^2_x\phi)^2-(\partial_x \phi)^3,
\label{s_cont}
\ee
where $\nabla=({\partial_{\bf z}},\partial_x)$ is the gradient in
$d+1$ dimensions. For the classical energy, as we mentioned above,
one has to substitute $d\to d-1$ in (\ref{s_cont}). The important
difference between the result for the continuum model
(\ref{s_cont}) and the lattice result (\ref{s6c}) is that the
power of the $k_c-k$ in the prefactor in (\ref{s_cont}) is smaller
than that in (\ref{s6c}). Moreover in $d=3$ the whole scaling
breaks down suggesting the the instanton action becomes
discontinuous at $d=3$ near the Mott transition.

We can evaluate the integral in (\ref{s_cont}) using the
variational ansatz. For simplicity we will take a separable
function
\be
\phi(z,x)=A(z){\tanh{\alpha x}\over \cosh{\beta x}},
\label{var}
\ee
where $\alpha$ and $\beta$ are the variational parameters and the
function $A(z)$ can be found solving remaining one-dimensional
problem. With this choice it is easy to show that in one and two
dimensions we obtain:
\be
\alpha_{1d}\approx 0.32,\;\beta_{1d}\approx 0.53,
\;s_{1d}\approx 73 (k_c-k)^{3\over 2}.
\label{act}
\ee
\be
\alpha_{2d}\approx 0.72,\;\beta_{1d}\approx 1.08,
\;s_{2d}\approx 67 (k_c-k)^{1\over 2}.
\label{act1}
\ee

In three dimensions this variational ansatz gives $S_3\to 0$,
which implies breaking of the scaling as $k_c\to k$. Indeed, the
power of $(k_c-k)$ in (\ref{s_cont}) becomes negative. However, it
is unphysical to expect any divergence near the point of
instability. This indicates that the original ansatz that the
longitudinal coordinate scales as $1/\sqrt{k_c-k}$ is not valid in
this case and the tunneling action becomes momentum independent as
$k\to k_c$. To see that this is indeed the case and to estimate
the actual value of $s$ in three dimensions we return to the the
action (\ref{act0}) without preforming rescaling (\ref{rescale})
and apply the variational ansatz (\ref{var}). Then as $k\to k_c$
we find:
\be
\alpha_{3d}\approx 0.53,\;\beta_{3d}\approx 0.8,
\;s_{3d}\approx  125.
\label{act2}
\ee

Using the mean field parameters $C$ and $\xi$ in (\ref{gl}) and
the results (\ref{act})-(\ref{act2}) we can rewrite the tunneling
action in the following form:
\beq
&&S_1\approx {5.7\over \xi^2}\left(1-\sqrt{3}\,p\,\xi\right)^{3/2},\label{s11}\\
&& S_2\approx {3.2\over \xi}
\left(1-\sqrt{3}\,p\,\xi\right)^{1/2},
\label{s22}\\
&& S_3\approx 4.3.\label{s33}.
\eeq
The equilibrium superfluid-insulator phase transition corresponds
to $\xi=\infty$. Notice that in one and two dimensions the
tunneling action is always small as long as $\xi\gg 1$ and $p$ is
close to the critical momentum. This implies that at small
currents the broadening of the nonequlibrium transition is very
large. This is consistent with earlier numerical
findings~\cite{pw} and recent experiments~\cite{chad}. In three
dimensions, as we argued above, the tunneling action is
discontinuous at the transition, therefore the mean field phase
boundary is defined very well. So at $d=3$ at zero temperature we
can accurately define a relatively sharp phase transition between
the current carrying superfluid and the insulator.

Quite similar considerations apply to the current decay due to
thermal fluctuations. The decay rate is determined by the ratio of
the activation energy $E_d$ and the temperature. In different
spatial dimensions we get:
\beq
&&E_1\approx 1.3 {JN\over
\xi^3}\left(1-\sqrt{3}\,p\,\xi\right)^{5\over
2},\\
&& E_2\approx 7.8 {JN\over
\xi^2}\left(1-\sqrt{3}\,p\,\xi\right)^{3\over
2},\label{e2}\\
&& E_3\approx 8.6 {JN\over \xi}
\left(1-\sqrt{3}\,p\,\xi\right)^{1\over 2}.
\label{e3}
\eeq
It is obvious that the thermal broadening is also strongest in one
dimension. However, contrary to the quantum case, even at three
dimensions close to the mean field transition the activation
barrier vanishes continuously. Only in four dimensions and above
we would be able to define a sharp phase boundary separating
current carrying superfluid and insulating phases at finite
temperature.

We would like to point out that the thermal decay is strongly
suppressed at low temperatures $T\ll JN/\xi$. Note that this
condition is also necessary in order to observe the equilibrium
superfluid-insulator transition and thus can be satisfied
experimentally. Another important point is that the action for the
quantum phase slip tunneling in one and two dimensions is never
large near the mean field critical current. This implies that the
quantum decay mechanism should be experimentally relevant at $d=1$
and $d=2$. This conclusion is similar to that we reached in the
previous section when we discussed current decay at small
interactions.

It is also possible to make some qualitative statements beyond the
mean field approximation in the vicinity of a quantum critical
point separating equilibrium superfluid and insulating phases.
Thus we can still expect that both the tunneling action and the
thermal activation barrier vanish at
\be
p\sim {1\over \xi}.
\ee
On the other hand, quite generally $1/\xi\sim \lambda^\nu$, where
$\nu$ is a critical exponent~\cite{sachdev_book} and $\lambda$ is
a tuning parameter across the transition, say deviation of the
interaction $U$ or hopping $J$ from the critical point. In the one
dimensional case $\nu=\infty$~[\onlinecite{Chaikin-Lubensky}]
(more precisely for the Kosterlits-Thouless transition
$\xi\propto\exp(b/\sqrt{\lambda})$, where $b$ is some constant).
In two and three dimensions the quantum critical point is
characterized by the universality class of classical $xy$-model in
one dimension higher and the corresponding critical exponents are
$\nu\approx 0.67$ at $d=2$ and $\nu=0.5$ at
$d=3$~[\onlinecite{Zinn-Justin}]. We see that in three dimensions
the mean field description gives the correct value of $\nu$. Also,
quite generally, we can argue that the action (\ref{gl}) and the
energy (\ref{gle}) remain valid near the quantum critical point,
but with constants $C$ and $C^\prime$ being renormalized:
\be
C\propto \xi^{d+z-4},\; C^\prime\propto \xi^{d-4}.
\ee
This in turn implies that near the quantum critical point
$\lambda\ll 1$ we get
\beq
&&S_d\approx \lambda^{\nu(3/2-z)}A_d
(B_d\lambda^\nu-p)^{5/2-d},\nonumber\\
&& E_d\approx JN A_d^\prime \lambda^{\nu/2}(B_d^\prime
\lambda^\nu-p)^{7/2-d},
\eeq
where $A_d$, $A_d^\prime$, $B_d$, and $B_d^\prime$ are
nonuniversal numbers. In the non-generic commensurate case, which
we are mostly interested in here, $z=1$. The scaling form above
agrees with the mean field results ($\nu=1/2$) obtained earlier.
Despite quantitative difference between the correct and mean field
scaling in one and two dimensions, the qualitative features of the
nonequilibrium phase transition discussed after (\ref{s22}) and
(\ref{e3}) remain intact.

\section{Dynamics of the decay. Influence of the confining potential.}
\label{sec3}

\subsection{Underdamped versus overdamped dynamics}
\begin{figure}[ht]
\includegraphics[width=4cm]{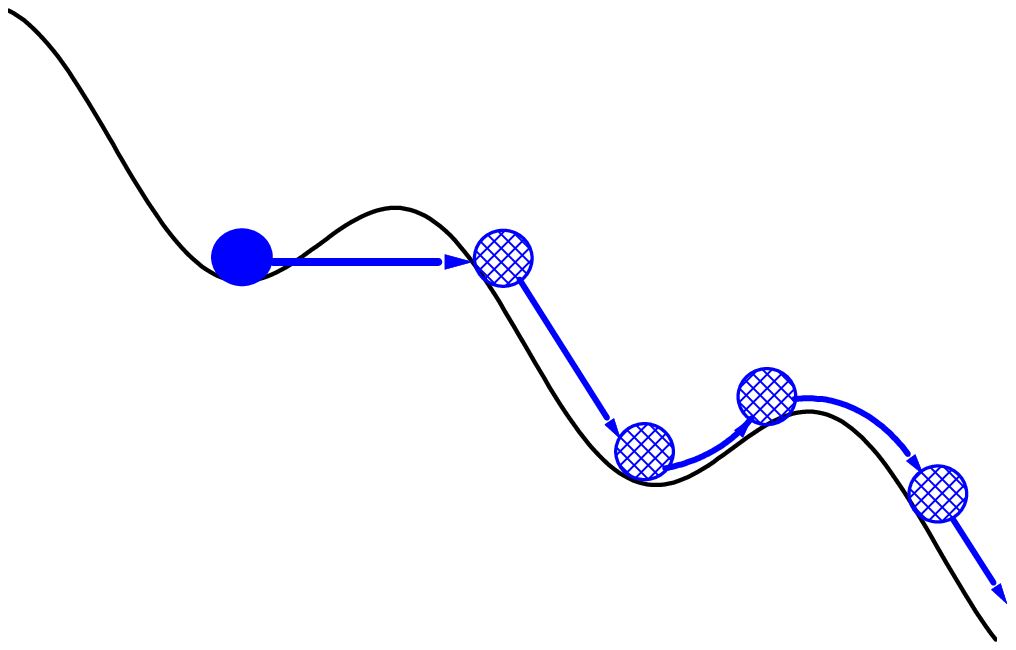}
\includegraphics[width=4cm]{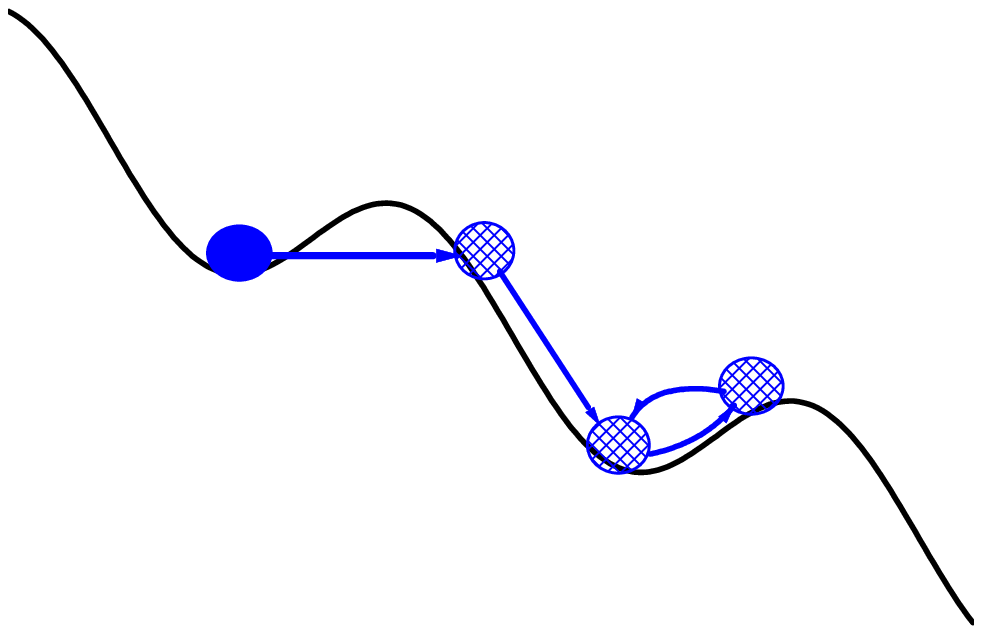}
\caption{Possible motion of a particle after a tunneling event in
a tilted washboard potential if the friction is small (left) and
if the friction is large (right). Instead of changing the
friction, one can vary the tilt. It is clear that reducing the
tilt is similar to increasing the friction.}
\label{fig_an}
\end{figure}

\begin{figure}[ht]
\includegraphics[width=8cm]{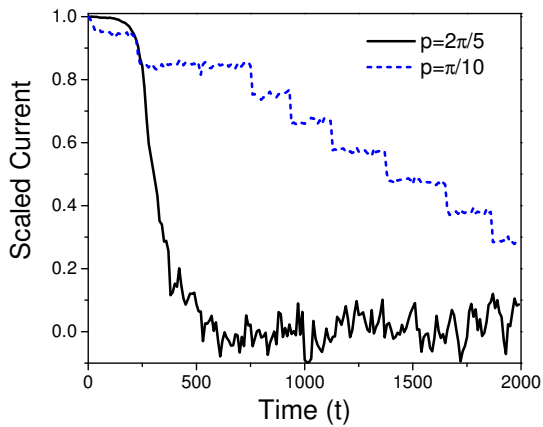}
\caption{Current (scaled to one at $t=0$) versus time for a one
dimensional periodic array of $200$ sites with two different
initial phase gradients. Here and in the following graphs time is
dimensionless. Its units are set by the inverse units of couplings
$J$ and $U$. The evolution is determined solving equations of
motion (\ref{gp1}) with constant hopping amplitude $J=1$ and
interaction increasing in time $U=0.01 \tanh 0.01 t$ for
$p=2\pi/5$ and $U=\tanh 0.01 t$ for $p=\pi/10$. To get the current
decay we add small fluctuations to the initial values of the
classical fields $\psi_j(t=0)$.}
\label{fig_damp}
\end{figure}

As we showed above, quantum or thermal fluctuations lead to the
broadening of the dynamical phase transition. However, this does
not imply that within a single experimental run a gradual current
decay will be necessarily detected as the system is slowly tuned
through the crossover region. The tunneling or the thermal
activation times define a probability of generating a single phase
slip. Once created the phase slip can either decay into phonon
(Bogoliubov's) modes and bring the system to a next metastable
minimum with a lower current, or this phase slip can trigger the
current decay in the whole system. This situation is analogous to
the motion of a particle on a tilted washboard potential with
friction (see Fig.~\ref{fig_an}). If the friction is large enough
(or the tilt is small) then the particle, after it tunnels, will
be stuck in the next minimum. On the other hand in the
frictionless case a single tunneling event will cause accelerated
motion of the particle through the whole lattice.
In a closed system these two regimes are well defined because the
damping of the phase slip motion comes from the internal degrees
of freedom, which are completely described by the equations of
motion.  To see which of the regimes is realized in our systems
within the classical thermal decay mechanism we numerically solve
Gross-Pitaveskii equations of motion for an array of condensates.
We start from a uniform current state. To have a current decay we
add small fluctuations to the initial values of the classical
fields. This is similar to starting from a thermal state.  Since
we cannot change the internal friction, instead we consider two
different tilts. In Fig.~\ref{fig_damp} we plot the computed
current versus time for a one-dimensional array of $M=200$ sites
with periodic boundary conditions. The initial state is the
noninteracting $U=0$ condensate with a given phase gradient $p$
(specifically $p=2\pi/5$ and $p=\pi/10$) and unit hopping.
It is clear from the figure that we have an overdamped case for
the smaller current case, where the phase slips occur one by one.
On the other hand for the larger current a single phase slip
generates immediate current decay in the whole sample and this
corresponds to the underdamped regime. We will not attempt here to
find the precise boundary between the two scenarios, since it is
not the purpose of our paper. We would like to emphasize that near
the mean-field instability the system is always underdamped, while
if the current decays at small $p$ the motion is overdamped. These
considerations agree with the experimental
observations~\cite{inguscio, chad}. We checked that the situation
is similar in other spatial dimensions. Thus, if the current
decays close to the mean field instability, then in a given
experiment one will observe a sharp transition from the superfluid
to the insulating regime. However, the precise point, where the
current decays will depend on the details of the experiment, for
example on the rate of change of external parameters like
tunneling and interaction or on the rate of change of the phase
gradient $p$ if the system is accelerated. On the other hand in
the absence of any fluctuations the transition is very sharp and
always occurs at $p=\pi/2$. We can also perform a similar analysis
close to the SF-IN transition. The qualitative result that close
to the modulational instability the phase-slip motion is
underdamped remains correct. However, we should stress that in one
and two dimensions broadening of the mean field transition is very
strong and the actual decay may occur very far from the critical
current. In this case an overdamped scenario should be realized.

Unfortunately we cannot simulate the dynamics of the decay due to
quantum tunneling. However, we would like to argue, that near the
critical current the fate of quantum and thermal phase slips is
identical. This is because the tunneling (activation) barrier is
very narrow and the classically allowed motion after the tunneling
event starts very close to the maximum of the barrier (see
Fig.~\ref{fig_an}).

We have to make another important remark that if the motion of
phase slips is underdamped then in a truly infinite system the
current state is always unstable. Indeed the probability of a
phase slip is proportional to the system size $M$. If it causes
the current decay in the whole system, then obviously a uniform
current state cannot exist. However, in finite size systems these
effects are not so crucial, because the decay probability depends
exponentially on the couplings and the current but only linearly
in the system size. So the phase diagram plotted in
Fig.~\ref{ph-dg} is quite robust to changes in $M$.

\subsection{Decay in a parabolic trap.}

If in addition to the optical lattice potential the condensate is
placed into a parabolic trap, then even at the classical
(Gross-Pitaevskii) level the condensate momentum is not a
conserved quantity. In a typical experiment, where the SF-IN
transition is probed, the lattice potential is slowly ramped up
resulting in hopping amplitude going down. In Appendix~\ref{seca3}
we show that in this case the amplitude of momentum oscillations
increases in time: $p\propto (1/J(t))^{1/4}$ (see Eq.~(\ref{14})).
If we ignore completely the effects of quantum fluctuations, then
for arbitrarily small initial displacement the condensate momentum
will ultimately exceed the critical value of $\pi/2$ and the
condensate motion will become unstable. If this happens before the
quantum fluctuations become significant, this effect will dominate
the SF-IN transition.
\begin{figure}[ht]
\includegraphics[width=8cm]{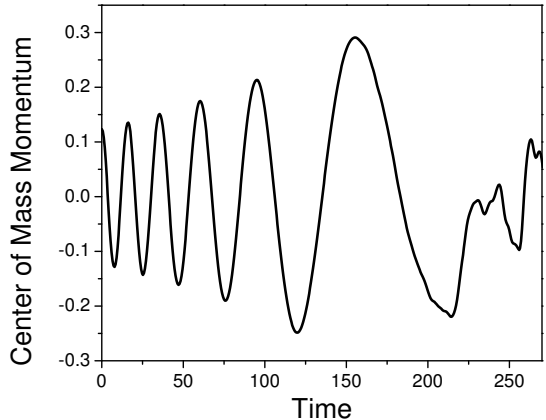}
\caption{Center of mass momentum (in the units of $\pi$ per
lattice site) versus time for a two-dimensional condensate in a
parabolic trap with hopping amplitude decreasing in time:
$J(t)=2.5\exp(-0.01 t)$. The other parameters are $U=1$, number of
atoms per central site $N_0=1.5$, strength of the confining
potential $k=0.02$.}
\label{fig_trap}
\end{figure}
In Fig.~\ref{fig_trap} we show the time evolution of the center of
mass momentum $p_{cm}=\sum_p p\, n_p/\sum_p n_p$ of the condensate
in a trap using Gutzwiller approximation. Here $n_p=\sum_{j,l}
\langle a_j^\dagger a_i\rangle \exp(ip\,(l-j))$ is the momentum
distribution function. The condensate initially in equilibrium is
given a small momentum boost. In agreement with our expectations
the amplitude of momentum oscillations grows in time. We note that
at the same time the condensate velocity $v(t)\propto \sqrt{J(t)
p(t)}$ decreases with time. This behavior continues until the
momentum exceeds a critical value, where the condensate motion
becomes chaotic.

One can avoid complications related to the nonconservation of the
amplitude of momentum oscillations by tuning the interaction strength
rather then the hopping amplitude. In this case one can directly
probe the boundary separating stable and unstable motion of
condensates at a given condensate momentum. Another important
feature, which distinguishes trapped condensates from homogeneous
systems is the spatial variation of the density. Thus if the
density profile is smooth enough, the condensate motion becomes
unstable first near the edges where $N\approx 1$. In the center of
the trap current decays at higher interactions. As we argued
earlier in this section, in homogeneous systems the motion of
phase slips is underdamped near the instability, i.e. a single
phase slip triggers current decay in the whole system. We can
expect the same to be true in a trap as long as the mean
occupation number in the central site $N_0$ remains close to
unity. On the other hand if $N_0\gg 1$ then it is intuitively
clear that phase slips occurring near the edges can not
destabilize the motion of the bulk of the condensate, which is
very far from the instability. To verify this reasoning
numerically we again employ Gutzwiller approximation.
\begin{figure}[h]
% \centering
\includegraphics[width=8cm]{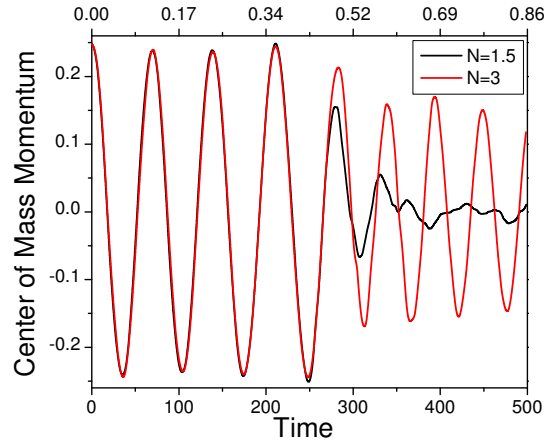}
\caption{Time dependence of the condensate momentum in a
two-dimensional harmonic trap with different filling factors per
central site. } \label{fig:trap}
\end{figure}
In Fig.~\ref{fig:trap} we plot the momentum oscillations of a
two-dimensional condensate versus time. We set the hopping
amplitude $J=1/4$ while increasing the interaction linearly in
time: $U(t)=0.01 t$. The simulations are performed on a lattice of
dimensions $120\times 60$ with global trapping potential
$V(j_x,j_y)=0.01(j_x^2+j_y^2)$. We consider two different filling
factors in the central site $N_0=1.5$ and $N_0=3$. It is obvious
that the onset of instability in both cases occurs at roughly the
same interaction strength (which is close to the uniform result at
filling factor $N=1$). However, while the motion becomes chaotic
very fast for $N_0=1.5$, there is a very gradual decay of momentum
oscillations at larger filling $N_0=3$. In agreement with our
expectations this indicates that an overdamped current decay is
realized in the latter case.

\section{Loss of coherence in the nonequlibrium phase transition.}
\label{sec4}

The superfluid to Mott insulator transition at equilibrium, is a
continuous quantum phase transition. As such, it is expected to be
reversible. That is, if we tune through the transition and then
back to the initial state slowly enough, we would recover
arbitrarily large fractions of the superfluid density~\cite{ap2}.

\begin{figure}
\includegraphics[width=8cm]{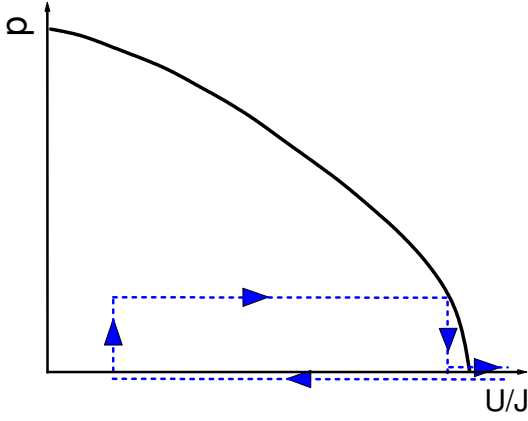}
\caption{Possible experimental sequence to observe a superfluid
insulator transition in a moving condensate.}
\label{loop}
\end{figure}

With finite current the situation is markedly different. We
envision the following experimental procedure. (1) The condensate
is boosted to small but finite velocity (dipole oscillation in a
trap). (2) An optical lattice is turned on adiabatically, until
the system passes the line of instability (See Fig.~\ref{loop}),
and then slowly turned off. Finally the atoms are released from
the trap, and their final momentum distribution is measured and
compared to the initial one.

The current carrying state has a finite energy, which is released
into the system when the current decays. Since the system is non
integrable, it is reasonable to assume that the energy associated
with the supercurrent, is released in the form of incoherent
excitations, which eventually thermalize. If now the system
parameters are changed sufficiently slowly, the subsequent
evolution of the system is adiabatic conserving the entropy of the
system. Thus, the entropy change following the current decay, may
be used to obtain the depletion of the superfluid in the final
state of the system. We would like to emphasize that these
assumptions apply if the parameters of the system change not too
fast, so that the current decays in a quasi static regime.
Otherwise the current decay and the entropy release are not
governed by the properties of the transition point and the actual
time dependent problem has to be solved.

The temperature of the equilibrium state reached following the current
decay, and before the system parameters had a chance to change appreciably
may be calculated by equating the energy of the superflow,
prior to its decay,
with the internal energy of the system in the new thermal equilibrium:
\be
\delta\e(p)={1\over M}\sum_{\bf q}{\omega_{q}\over \mathrm
e^{\beta\omega_q}-1}.
\label{beta}
\ee
The low energy excitations in the superfluid state are linearly dispersing
sound modes with $\w_\bk\approx c k$.
Substituting this in (\ref{beta}), we can solve for the temperature to obtain:
\be
T=A_d c^{d\over d+1} \d\e(p)^{1\over d+1}
\ee
where
\be
A_d=\left(\frac{(2\pi)^d}{\W_d d!\zeta(1+d)}\right)^{1\over d+1},
\ee
$\W_d$ is the surface area of the $d$-dimensional unit sphere, and
$\zeta$ stands for the Riemann zeta function. Accordingly, the
entropy of this thermal state is given by:
\be
\mathcal S={\d\e(p)\over T}-\sum_{\bf q}\log
\left(1-\mathrm e^{-\beta\omega_q}\right)\approx
A_d^{-1}\left({\delta\e\over c}\right)^{d\over d+1}
{d+1\over d}.
\label{S1}
\ee
As argued above, we can use this entropy to calculate the
condensate depletion in the final state.

Let us apply this procedure assuming that the current decays via
the instability in the vicinity of the Mott transition. Though we
do the calculation for all dimensions, one should note that such a
scenario is particularly relevant for a three dimensional optical
lattice. According to section \ref{sec3}, in lower dimensions the
current will most probably have decayed due to fluctuations before
reaching the instability.

The energy per site of a state near the Mott transition according
to the Ginsburg-Landau model, (see Appendix~\ref{sec:ax}) is given
by:
\be
\e=\frac{JN}{2 d u V}\int d^d
x\left[|\nabla\psi|^2-\xi^{-2}|\psi|^2+\half|\psi|^4\right]
\ee
Substituting the field corresponding to the current carrying
state: $\psi=\sqrt{\xi^{-2}-p^2}e^{i p x}$, we obtain
\be
\d\e=\e(p)-\e(0)=\frac{JN}{2 d u}(\xi^{-2}-\half p^2)p^2.
\ee
Recall that our proposed experiment maintains constant $p$ and
changes the dimensionless interaction, and hence also $\xi$, by
increasing the lattice intensity. The current decays when the
instability is reached, i.e. when $\xi^{-2}=\xi_c^{-2}=3 p^2$, at
which point the energy per site is:
\be
\d\e\approx \frac{JN}{2 d}{5\over 2} p^4.
\ee
Thus the energy released following decay via the instability at
phase gradient $p$ is $\propto p^4$. Using (\ref{S1}) and the
sound velocity near the transition $c=2JN\sqrt{2 d}$ we get the
increase of entropy per site:
\be
S=\frac{d+1}{d\cdot A_d}
\left(\frac{5}{4(2d)^{3/2}}\right)^{d\over d+1} p^{\frac{4
d}{d+1}}
\ee

In three dimensions, this gives $S_{3d}\approx 0.16 p^3$. With
such a small increase of entropy we anticipate that the
irreversibility, as manifest in the unrecovered condensate
fraction, would also be small, for low initial currents. Perhaps
the simplest way to see this is to slowly reduce the lattice
intensity until the atoms are very weakly interacting before
releasing to measure the momentum distribution. In this case the
elementary excitations have a quadratic dispersion $\w_\bk=\a
k^2$. In general this assumption is not necessary and one can use
the full Bogoliubov spectrum. However, qualitatively the result
remains the same. A nice feature of the quadratic dispersion, is
that the thermal depletion is simply related to the entropy, which
is given by:
\be
S=\left(\frac{d+2}{d}\right){E\over T}
=\frac{\W_d}{(2\pi)^d}\left(\frac{d+2}{d}\right)
\left(\frac{T}{\a}\right)^{d/2}I_d
\ee
Where
\be
I_d=\int_0^{\infty} dx \frac{x^{d+1}}{e^{x^2}-1}.
\ee
The thermal depletion on the other hand is
\bea
n_T&=&\frac{1}{V}\sum_\bq \frac{1}{e^{\a q^2/T-1}}\nn\\
&=&\left(\frac{d}{2+d}\right)\frac{S}{I_d}
\int_{x_0}^\infty dx x^{d-1}\frac{1}{e^{x^2}-1},
\eea
where $x_0$ depends on the small momentum cutoff determined
by the system size (i.e. $x_0\sim  L\sqrt{\a/T}$).
In three dimensions the last integral is convergent and we get:
\be
n_T^{3d}=\frac{2\zeta(3/2)}{5\zeta(5/2)} S\approx 0.78 S
\label{n35}
\ee
Thus, in three dimensions the number of excited quasiparticles, or
condensate depletion, is a direct measure of the entropy

In one and two dimensions on the other hand, the integral has an
infrared divergence. Of course this is simply a restatement of the
well known fact that a true condensate in free space at finite
temperature does not exist below three dimensions. In practice,
the divergence is cut off by the system size and one can define
the notion of a quasicondensate. In two dimensions where the
divergence is only logarithmic:
\be
n_T^{2d}\approx \frac{6}{\pi^2} S\log (L\sqrt{S})
\ee
And finally in one dimension
\be
n_T^{1d}\approx 0.04 S^2 L
\ee
Instead of the system size the cutoff may come from the finite
momentum resolution ($\Delta p$) of the experimental apparatus. In
general the momentum cutoff will be determined by the minimum of
$L$ and $1/\Delta p$. We see that below three dimensions
condensate depletion due to thermal decay at small currents is
more pronounced and easier to detect than at $d=3$.

The situation is different if the current decays at smaller
interaction strengths (smaller lattice intensity) at large
currents. The energy of the current state may then be calculated
from the Gross-Pitaevskii energy functional:
\be
E=-J\sum_\av{ij}(\psi^\star_i\psi_j+\psi^\star_j\psi_i)+{U\over
2}\sum_i|\psi_i|^4
\ee
Substituting $\sqrt{N}e^{i p x_j}$ for $\psi_j$ we find
$\d\e=4JN\sin^2(p/2)$. Together with Eq. (\ref{S1}), and using the
sound velocity $c=\sqrt{2UJN}$ we get the entropy increase per
site:
\beq
&&S_{1d}\approx 2.4\left({JN\over U}\right)^{1/4}\sin{p\over 2},\label{ss1}\\
&&S_{2d}\approx 2.2 \left({JN\over
U}\right)^{1/3}\sin^{4/3}{p\over 2},\label{ss2}\\
&&S_{3d}\approx 2.2 \left({JN\over
U}\right)^{3/8}\sin^{3/2}{p\over 2}.\label{ss3}
\eeq
Evidently the irreversibility is stronger and easier to detect if
the current decays at smaller interaction strengths.

\section{Exact results in small systems.}
\label{sec:exact}

In this section we present results of exact dynamics in small
one-dimensional systems. We will assume that we have a periodic
one-dimensional array of $M$ sites containing $N$ particles. At
$t=0$ we assume that interactions are absent and the
system is placed in the uniform current state, described
by the wave function:
\be
|\psi\rangle={1\over \sqrt{N!}}\left({\sum_{j=0}^{M-1}
a_j^\dagger\mathrm e^{ipj}\over \sqrt{M}}\right)^{N}|0\rangle,
\ee
\begin{figure}[ht]
\includegraphics[width=8cm]{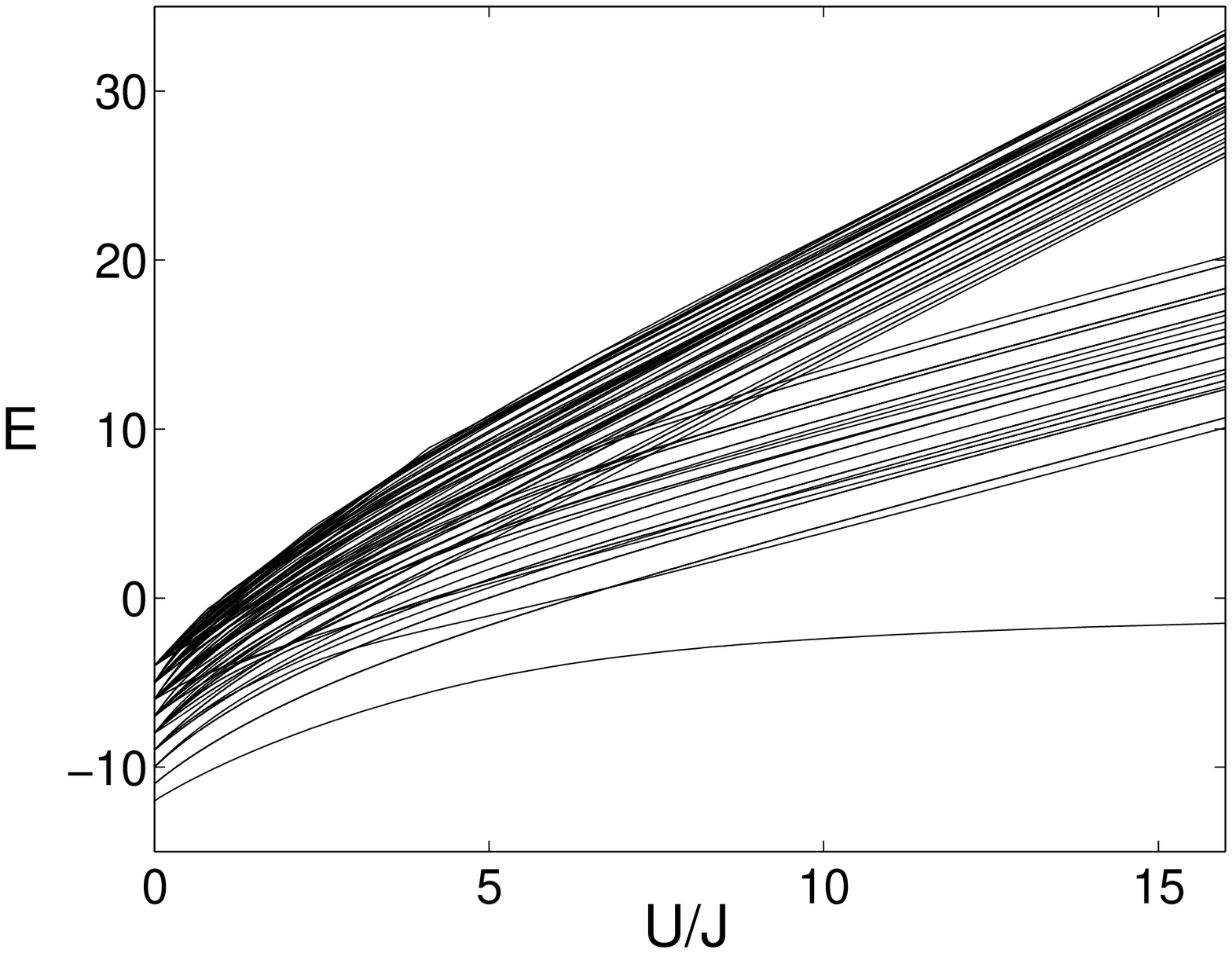}
\includegraphics[width=8cm]{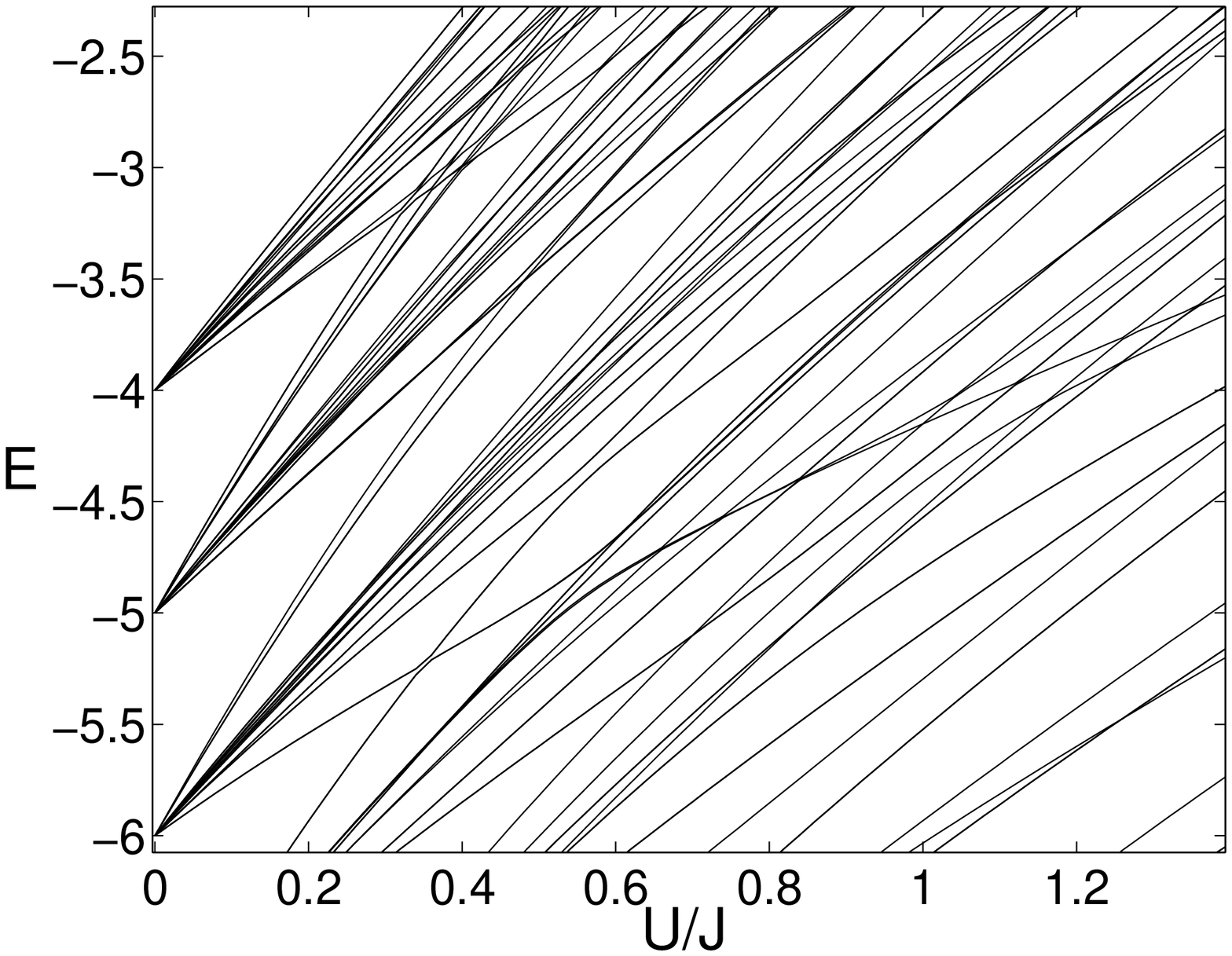}
\caption{First one hundred energy levels of the spectrum in a
periodic array of six sites containing six particles (top) and
zoom of the energy spectrum around the current carrying state with
$p=\pi/6$ (bottom). It is clear that while the ground state is
well separated from the continuum for all interaction strengths,
the excited many body states are completely mixed. So it is nearly
impossible to be in the adiabatic regime unless the system is
initially in the ground state.}
\label{fig:es}
\end{figure}
where $|0\rangle$ is the no-particle vacuum. This state is an
exact eigenstate of the noninteracting system. Now we slowly turn
on interactions driving the system closer to the insulating regime
and then gradually turn them off. The latter step is necessary to
check whether we have reversible dynamics or irreversible current
decay. Although in real experiments it is rather tunneling not the
interaction which is changed in time, this does not make any
qualitative difference in uniform systems. If the interaction is
ramped up infinitesimally slowly, then any finite system will
remain within a particular energy eigen-state and the evolution
will be always reversible. However, the energy splitting between
the many body levels decreases exponentially with the system size and
the number of particles. So practically even in relatively small
systems, one can study dynamics, which is slow with respect to the
characteristic time scales (like period of Josephson
oscillations), but very fast with respect to the inverse many body
energy spacing. To make this point more transparent we plot in
Fig.~\ref{fig:es} the energy spectrum versus $U/J$ for a
particular system of a periodic array consisting of six sites and
containing six particles. The size of the Hilbert space here is
already quite big: $\mathcal N=11!/(6! 5!)=462$. It is obvious
from the figure that while the ground state is well separated from
the continuum at all interaction strengths, the excited states
experience many level crossings. So it is nearly impossible to
be in or close to the adiabatic regime unless the system is in
the ground state.

Let us assume that the Hamitonian is described by (\ref{hamilt}),
where the interaction strength changes in time according to
\be
U(t)=U_0\tanh\delta t\,\tanh \delta(T-t), \label{int_t}
\ee
where $\delta$ is the adiabaticity parameter, $T$ is the duration
of the time evolution and $U_0$ is a prefactor setting the energy
scale. We also assume that the hopping $J$ is equal to unity and
does not change in time.
\begin{figure}[ht]
\centering
\includegraphics[width=8cm]{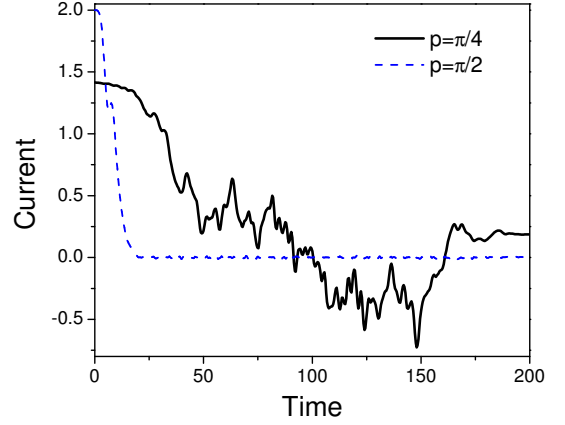}
\caption{Dependence of current on time for the system of eight
sites with two particles per site. The hopping amplitude is equal
to unity and is independent of time. The interaction changes in
time according to Eq.~(\ref{int_t}) with $U_0$=2, $\delta=0.02$,
and $T=200$.} \label{cur_t}
\end{figure}

In Fig.~\ref{cur_t} we plot the current versus time for the array
of eight sites with two particles per site. The two curves
correspond to initial phase gradient of $\pi/4$ and $\pi/2$ per bond.
The
smaller current decays when the interaction becomes sufficiently
large, in agreement with that this state is metastable, while the
$\pi/2$ state decays almost instantly since it is classically
unstable. In both cases the decay is clearly completely
irreversible.
\begin{figure}[ht]
\centering
\includegraphics[width=7.5cm, angle=90]{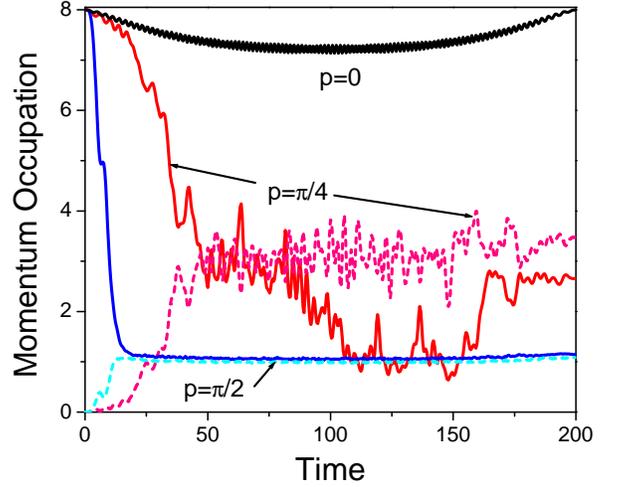}
\caption{Occupation of momentum states as a function of time for
different current carrying states. Solid lines correspond to the
occupation of momentum $p$ equal to the initial phase gradient in
the system. The dashed lines are the occupations of the
zero-momentum state. The parameters are the same as in
Fig.~\ref{cur_t}.} \label{md_t}
\end{figure}
To make the final point more transparent we also plot momentum
distribution as a function of time for the same parameters in
Fig.~\ref{md_t}. The curve corresponding to a zero current state
shows reversible behavior, suggesting that the interaction indeed
changes slowly enough so that the system is in the adiabatic
limit. Note, that even at the peak interaction strength, the
system is still far from the insulating state, as evident
from the very small depletion of the $p=0$ state.
Nevertheless, because the smallest phase difference per site
achievable in an array of size eight is still quite large
($p=\pi/4$), this interaction is sufficient to drive
current decay. Indeed the curve corresponding to initial
phase gradient of $\pi/4$
clearly displays metastable behavior, with current decaying
after some delay. In the new steady state reached, occupation of
zero momentum builds up suggesting that it
corresponds to a thermal distribution with no macroscopic
current, but with some residual phase coherence.
The state with phase gradient $p=\pi/2$, by contrast,
seems to decay into a high temperature state,
without visible phase correlations. Indeed, occupation
of momenta zero and $\pi/2$ are equal to unity
as expected if the phases were completely random.
It is peculiar that
there are only very weak fluctuations of the momentum
distribution in this state. This should be contrasted with the
rather large fluctuations seen following decay of the
$\pi/4$ current state.

Finally Condensates sustaining phase gradients
$p>\pi/2$ are classically unstable, and are expected to
decay rapidly, resulting in even higher
temperature than for the $\pi/2$ state. The $p=\pi$ state
is an interesting exception to this rule. As
argued in~\cite{ap1,psg}, this state evolves into a
maximally entangled Schr\"odinger cat-like state, which is robust
to weak perturbations in small systems.
\begin{figure}[ht]
\centering
\includegraphics[width=7.5cm, angle=90]{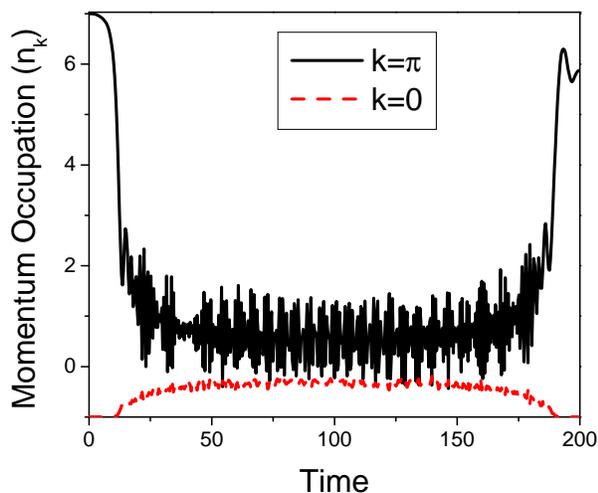}
\caption{Occupations of the $k=0$ and $k=\pi$ states for the state
with initial phase gradient
$p=\pi$. Here the interaction changes in time according to
`(\ref{int_t}) with $U_0$=1, $\delta=0.02$, and $T=200$.}
\label{cat}
\end{figure}
Physically this happens because the $\pi$ state in the
noninteracting case is the most excited state in the system. In
the absence of energy relaxation the system remains in the most
excited state under slow perturbations in the same way as it does
in the ground state. And therefore we can expect reversible
behavior of the phase coherence. We plot the corresponding
momentum distribution for this state in Fig.~\ref{cat}. Although
for the intermediate times the evolution looks completely chaotic,
once the interaction is reduced back to zero the momentum
occupancy at $k=\pi$ reaches almost the maximum possible initial
value suggesting only a small amount of excitations in the system.

Unfortunately doing exact numerical simulations we are quite
limited by the total system sizes and the number of particles.
Also we can address only one-dimensional systems. Therefore we
can not directly test the phase diagram and decay rates derived in
the previous sections. Nevertheless, we would like to point that
the numerical results are in excellent qualitative agreement with
our predictions.

\section{Summary and experimental implications.}
\label{sec:fin}

In summary we emphasize two important predictions of this work. In
Sec.~\ref{sec1} we derived a mean field phase diagram for the
stability of a moving condensate. In Sec.~\ref{sec2} we
investigated the effect of quantum and thermal fluctuations, which
lead to broadening of the nonequilibrium transition.

We showed that the mean field transition interpolates between the
classical modulational instability, which occurs at phase gradient
$p=\pi/2$ deep in the superfluid regime and the equilibrium
($p=0$) transition to the Mott insulating phase at strong
interactions. The dynamical transition is of first order (i.e. irreversible)
at any nonzero current, contrary to the second order transition at
equilibrium. Thus, if one starts from a current state and slowly
drives the system towards the insulating regime, e.g. by ramping
up the lattice potential, then after crossing the transition
boundary the current decays irreversibly, releasing the energy of
the coherent motion into heat. Plotting the location of the
nonequilibrium transition as a function of the current and
extrapolating the curve into the static regime $p=0$ is a way to
accurately determine the position of the equilibrium SF-IN
transition.

The mean field theory does not take into account quantum tunneling
and thermal activation of phase slips. These induce decay of
supercurrent, even before the classical equations of motion become
unstable. We calculated the asymptotic decay rates near the mean
field instability in two regimes: (i) deep in the superfluid regime
and (ii) close to the equilibrium Mott transition.

In a three dimensional optical lattice the broadening of the
transition due to these effects is found to be small in all cases.
In particular we find a discontinuity in the current decay rate
across the mean field transition for small currents at zero
temperature (close to the equilibrium Mott transition). Thus the
dynamical transition survives the effect of quantum fluctuations
in this case. We predict that a sharp dynamical
superfluid-insulator transition would be seen at small currents at
a critical interaction strength (or lattice depth).

In one and two dimensions, on the other hand, quantum and thermal
phase slips lead to substantial broadening of the transition,
especially when the average site occupation $N\sim 1$. Then we
expect the current to decay well before the dynamical instability
is reached. Indeed, in a recent experiment~\cite{chad} strong
damping was detected at currents much smaller than that given by
the Gross-Pitaevskii modulational instability. In addition the
observed dependence of the damping rate on the lattice depth
potential was very smooth. This is in line with our prediction of
a the large broadening of the mean-field transition by quantum
fluctuations and should be contrasted with the Gross-Pitaveskii
predictions of a sharp transition. It is also consistent, with
earlier numerical results by one of us~\cite{pw}.

The experimental results \cite{chad}, do not by themselves prove
that quantum rather than thermal fluctuations are responsible for
the observed damping. Here we estimate the crossover temperature
from thermal to quantum dominated decay to be of order the
Josephson frequency $T^\star\approx \sqrt{JUN}/k_B$. The
experiment is therefore likely to be dominated by quantum phase
slips. To verify this conclusions, one could measure e.g. the
damping as a function of temperature and observe a saturation of
the rate around $T^\star$.

One perhaps surprising experimental observation was that in the
overdamped regime (i.e. at high optical lattice potential) the
condensate was  essentially localized in a tilted lattice, while
it still exhibited sharp coherence peaks~\cite{chad}. A possible
explanation is the effect of the inhomogeneous density in a harmonic
trap, which we discussed in Sec.~\ref{sec3}. Indeed, in the
overdamped regime one expects no suppression of phase slip
tunneling at the edges of the condensate, where $N\approx 1$. This
implies that the phase at the edges fluctuates very wildly and the
edge atoms are localized. On the other hand,
in the middle of the condensate, where the mean number of
particles per site is larger, the system is far from
the mean field transition and phase slips are relatively
costly. As a result, the edges of the condensate create an
effective potential barrier stopping the motion of the rest of the
system, which retains phase coherence.

Our results are consistent with another recent experiment,
where the superfluid decay was measured as a function of the
condensate velocity in a one-dimensional optical
lattice~\cite{inguscio1}. There the average number of bosons per
site was large and quantum fluctuations negligible. At low
temperature a dynamical localization transition consistent with
Gross-Pitaveskii predictions was observed. However, at relatively
high temperatures, the motion became unstable at much lower
quasi-momentum. This observation qualitatively agrees with the
decay mechanism due to thermal phase-slips considered in the
present work.

We also presented exact numerical simulation of small
one-dimensional systems. These were in qualitative agreement with
the physical picture discussed in the paper. For example we
demonstrated that at nonzero current the transition is
irreversible. We also find that in a periodic chain of eight
sites, the current state with $p=\pi/4$ decays only at some finite
interaction strength, while at $p=\pi/2$ the decay occurs almost
instantly. An important exception is the case with $p=\pi$, where
the evolution is reversible (see also Ref.~[\onlinecite{ap2}]). A
quantitative comparison of the exact numerical results with our
predictions is not possible because of strong finite size effects
in the exact simulations.

{\em Note added ~--~} After this paper has been submitted two
preprints appeared~\cite{rey,davis}, which address the experiment
by Fertig \etal ~\cite{chad}. There, the damping was attributed to
single particle Bloch oscillations in the free fermion
representation of the bosons in the limit of strong interactions.
This effect can also be understood in the Boson language. If the
number of particles in the trap is small, the system reaches the
impenetrable boson regime while not yet insulating. Indeed one can
easily transfer a boson from a filled to an empty site near the
edge of the system. If the tunneling amplitude ($J$) is larger
than the single particle energy near the edge of the cloud
$J>kN_t^2/8$ then the created hole is delocalized through the
whole system and the state is not insulating. Here $kj^2/2$ is the
confining potential of the trap and $N_t$ is the total number of
particles, which equals the system size in the fermionized regime.
The hard core constraint in turn requires that $U\gg J$. Therefore
if $U\gg kN_t^2$ and we gradually decrease $J$ then indeed the
system first goes to fermionized delocalized regime $U\gg J\gg
kN_t^2$ and only if $J$ is decreased further it becomes localized.
On the other hand if the total number of bosons is large $U\ll
kN_t^2$, which is the case close to the thermodynamic limit, then
the edge excitations in the fermionized regime are always
localized and thus unimportant for the macroscopic properties of
the system.

We emphasize that if the first (i.e. small particle number
scenario) is realized, then after the trap minimum is displaced,
the particles will essentially undergo Bloch oscillations with
different frequencies resulting in a damped center of mass
motion~\cite{rey}. In this scenario there is no real energy
relaxation of the center of mass and it will saturate at a
displaced position~\cite{davis}. In the second case $U\ll kN_t^2$
the system undergoes a Mott transition when the impenetrable
regime is reached and the edge bosonic excitations can result only
in a tiny center of mass displacement vanishing in the
thermodynamic limit. The damping prior to the Mott transition in
this case occurs via the mechanisms discussed in this paper. I.e.,
the current decay is irreversible and results in energy relaxation
of the center of mass motion, which will eventually slide to the
minimum of the trap. This seems to be the case realized in the
experiment of Ref.~[\onlinecite{chad}].

We also mention that the single particle Bloch physics will
dominate the decay mechanisms studied here if the Bloch
oscillation, which frequency is equal to the single particle
energy separation between the nearest sites due to external
potential, is longer than the Josephson oscillation. This
condition is satisfied by the experimental systems of
interest~\cite{inguscio,chad}.

\acknowledgements

We would like to acknowledge D.-W. Wang for helpful discussions
and collaboration on related work. This work was supported by
the NSF (grants DMR-01328074, DMR-0233773, DMR-0231631,
DMR-0213805, PHY-0134776), the Sloan and the Packard Foundations,
and by Harvard-MIT CUA.

\appendix
\section{Derivation of a Ginzburg-Landau
action near the suprfluid insulator transition}
\label{sec:ax}

A derivation of a Ginzburg-Landau action near the superfluid
insulator transition was outlined in Ref.  ~[\onlinecite{ehud}].
For convenience we present the full derivation in this appendix.

Let us choose the energy of a single site with integer $N$ atoms,
as the zero of energy. Then the Hamiltonian of the boson Hubbard
model~(\ref{bh}) assumes the form:
\be
H=-J\sum_{\av{ij}}(a\yd_ia\nd_j+H.c.)+\sum_i
\frac{U}{2}(n_i-N)^2-\mu(n_i-N)
\label{BHM2}
\ee
Close to the superfluid insulator transition, the particle number fluctuation
is small. It is then possible to consider a subspace
allowing only occupations of $N-1$, $N$ and $N+1$ atoms per site.
This reduced Hilbert space is conveniently described by a (over-complete)
basis of product states:
\begin{widetext}
 \be
\ket{\W} =\prod_j\left[\cos (\t_j/2 )
\ket{N}_j+e^{i\h_j}\sin (\t_j/2 )
\left(e^{i\f_j}\sin (\x/2 ) \ket{N+1}_j
+e^{-i\f_j}\cos (\x/2 )\ket{N-1}_j\right)\right] .
\label{Omega}
\ee
We shall use these states to construct a path integral
for the evolution operator.
The derivation can be carried out for arbitrary $N$, but for simplicity
of presentation we take $N>>1$.
The first step is to prove a resolution of identity:
\be
\int_0^{\pi}d\t\int_0^{\pi}d\x\int_0^{2\pi}d\f\int_{-\pi/2}^{\pi/2}d\eta
{\cal M}(\W)\ket{\W}\bra{\W}={\cal I} ,\label{ResI-0}
\ee
We shall now find a suitable integration measure ${\cal M}(\t)$,
which is a function of $\t$
only. Substituting ${\cal M}(\t)$ in (\ref{ResI-0}),
we can integrate over $\h$, $\f$ and $\x$, which
kills off the cross terms so that (\ref{ResI-0})
reduces to:
\be
{\cal I}= 2\pi^3\int_0^{\pi}d\t{\cal
M}(\t)\Bigg\{\cos^2\frac{\t}{2}
\ket{N}\bra{N}+\half\sin^2\frac{\t}{2}\Big (\ket{N+1}\bra{N+1}
+\ket{N-1}\bra{N-1}\Big)\Bigg\}
\ee
\end{widetext}
The measure ${\cal M}(\t)$ must enforce the identity between
the diagonal matrix elements, so that:
\be
\int_0^{\pi}d\t{\cal
M}(\t)\left(\cos^2\frac{\t}{2}-\half\sin^2\frac{\t}{2}\right)=0
\ee
or equivalently
\be
\int_0^{\pi}d\t{\cal M}(\t)(1+3\cos\t)= \int_{-1}^1
dy\frac{{\cal M}(y)(1+3y)}{\sqrt{1-y^2}}=0,
\ee
where $y=\cos \theta$. This requirement is satisfied by ${\cal
M}(\t)=C \cos\t(3\cos\t-1)$, since it ensures that the integrand
is an antisymmetric function of $\cos\t$. The constant
$C=\pi^{-4}$ is determined from normalization. Since we are
interested in the vicinity of the transition at $\t=0$, where the
measure ${\cal M}$ changes slowly, it is safe to replace it with a
constant.

It is also straightforward to calculate the Berry phase
\be
\bra{\W}\frac{d}{dt}\ket{\W}=i\sum_j \sin^2(\t_j/2)(\dot{\h_j}
-\cos\x_j\dot{\f_j})\equiv -i\Upsilon(t).
\label{dWdt}
 \ee
Equations (\ref{ResI-0})  and (\ref{dWdt}) are the necessary
ingredients for the  path integral representation of the evolution
operator:
\bea
{\cal U}(t)=\int{\mathcal D} \W \exp\Big\{i\int_0^t
dt'\big[\Upsilon(t')-\cH (t')\big]\Big\},
\label{U}
\eea
where the classical Hamiltonian is given by
\begin{widetext}
\bea
\cH=\bra{\W}H\ket{\W}=\sum_i\sin^2(\t_i/2)(\frac{U}{2}-\mu\cos\x_i)
&-&2JN\sum_{\av{ij}}\rho_i\rho_j\Big[c_ic_j\cos(\h_j-\h_i+\f_i-\f_j)
+s_i s_j \cos(\h_j-\h_i+\f_j-\f_i)\nn\\
&&+ c_i s_j\cos(\h_i+\h_j+\f_j-\f_i)+s_i
c_j\cos(\h_i+\h_j+\f_i-\f_j)\Big].
\label{cH}
\eea
Here we introduced the notations $c_i\equiv\cos(\x_i/2)$,
$s_i\equiv\sin(\x_i/2)$ and $\rho_i\equiv\half\sin\t_i$. It is
important to note, that the dynamics defined by (\ref{dWdt}) and
(\ref{U}),  consists of two pairs of conjugate variables. The
average offset from integer density is conjugate to the phase
$\f$, while the second moment (i.e. the number fluctuation) is
conjugate to $\h$.

At fillings close to integer, the minimum classical energy is
reached in a uniform state with $\x$ close to $\pi/2$ and $\h=0$.
We therefore expand the action to leading order in
$\s_i=\pi/2-\x_i$ and $\h_i$. In addition we expand up to quartic
order in $\rho$, and anticipating a diverging length scale at the
transition, take the continuum limit of the action via a gradient
expansion
\be
S\approx \int dt' d^d x\left[\rho^2\s(\dot\f+\mu)-\rho^2\dot\h
-2JN\left((\nabla\rho)^2+\rho^2(\nabla\f)^2\right)
+4JNd(1-u)\rho^2-4JNd\cdot u\rho^4-2JNd\cdot\rho^2(2\h^2+\s^2/2)\right],
\label{Scont}
\ee
where $u=U/8JNd$. We can now integrate over the gaussian
fields $\h$ and $\s$ to
obtain a Ginzburg-Landau (GL) action:
\bea
S&=&\int dt' d^d x\Big[\frac{1}{4JNd}
\left(\dot\rho^2+\rho^2(\dot\f+\mu)\right)
-2JN\left((\nabla\rho)^2+\rho^2(\nabla\f)^2\right)-4JNd(1-u)\rho^2
+ 4JNd\cdot u\rho^4\Big]\nn\\
&=&{1\over 4JNd}\int dt' d^d x\Big[
|(\partial_t+i\mu)\psi|^2 -(2JN)^2 2d|\nabla\psi|^2+(4JNd)^2 (1-u)
|\psi|^2-(4JNd)^2 u|\psi|^4\Big]
\eea
\end{widetext}
where $\psi\equiv \rho e^{i\f}$. Note, that in Eq.~(\ref{Scont})
we left out terms of the form $\rho^2(\nabla\h)^2$ and
$\rho^2(\nabla\s)^2$. After integrating over $\h$ and $\s$, these
would lead to irrelevant high order derivatives in the GL action.
We can identify a sound velocity $c=2JN\sqrt{2d}$, where the
lattice constant is set to $l=1$. If we now make the
transformation $t\to c t$ (i.e measure time in units of $l/c$),
and $\psi\to \psi \sqrt{4du}$, the GL action assumes the form:
\be
S=\frac{1}{\a}\int dt
d^dx|(\partial_t+i\mu)\psi|^2-|\nabla\psi|^2
+r|\psi|^2-\half|\psi|^4,
\label{action0}
\ee
where $r=2d(1-u)$ and $\a=2u(2d)^{3/2}$.
In the superfluid phase $r=\xi^{-2}$, where
$\xi$ is the mean field coherence length. Note that the action
is Lorentz invariant only at commensurate filling, where by our
choice of the zero of energy for (\ref{BHM2}), $\mu=0$. This
is due to the particle hole symmetry,
which is present only at commensurate filling.

From the action (\ref{action0}), we can find the deviation from
commensurate filling, given by the conserved charge:
\be
Q=\frac{\d S}{\d \mu(t)}=\frac{1}{\a i}\int
d^dx \left[\psi^\star (\partial_t+i\mu)\psi
-\psi(\partial_t-i\mu)\psi^\star\right].
\label{charge}
\ee
In (\ref{BHM2}),
we chose the zero of energy such that the chemical potential
is $\mu=0$ at commensurate
filling $N$. Indeed if we substitute $\mu=0$ in (\ref{charge})
and a time independent order parameter we get $Q=0$ as required.
However, we note that the time derivative always appears in a gauge
invariant combination with the chemical potential. Therefore the
action (\ref{action0}) and the ''charge''
(\ref{charge}), dot not depend on the particular choice of the zero
of energy. In particular for calculating the dynamics
it would prove convenient to use a different gauge, applying
the transformation $\psi\to \psi e^{i\phi(t)}$,
$\mu\to\mu-\dot\phi$ with $\phi=\mu t$. This eliminates $\mu$
from the action, at the expense of imposing on the order parameter
an aditional  time dependent phase. The two gauges coincide
at the commensurate point where $\mu=0$ and there is no
time dependent phase. At incommensurate filling, though the action
seems Lorentz invariant in the new gauge,
the physics is clearly not, due to the imposed time dependent phase

We also note that in any gauge we can trace back
the density parameter $\s=\pi/2-\x$,
appearing in the Gutzwiller states (\ref{Omega}).
In mean field theory, integrating
out $\s$ in (\ref{Scont}), simply enforces the identity
$\s=\sqrt{8/ d}(\dot\f+\mu)$.
A small incommensurate filling is then given by $\s\rho^2$.

The Euler-Lagrange equations derived from the action in
the new gauge (where $\mu$ is eliminated from the action),
are given by Eq.~(\ref{gpr}), which we reproduce here for
completeness
\be
{\partial^2\psi\over \partial t^2}={\bf
\nabla}^2\psi+r\psi-|\psi|^2\psi,
\label{gprA}
\ee
In this gauge
the density offset from integer filling is
set exclusively by the initial conditions
for $\psi$, and $\dot\psi$ and given by
\be
Q=\frac{1}{\a i}\int d^d x(\psi^\star\dot\psi-\psi\dot\psi^\star).
\label{n2}
\ee
It is easily verified that (\ref{n2}), is a conserved quantity in
the equations of motion (\ref{gprA}). This gauge choice, is thus
analogous to the canonical ensemble, where the particle number is
fixed and is automatically conserved by the dynamics at all
later times.

Before concluding this appendix let us make several notes. First, to
obtain the action (\ref{gl}) from Eq.~(\ref{action0}), one has to
rotate to imaginary time $t= ix_0$, rescale length $\bx \to \xi
\bx$ (i.e measure length in units of the coherence length), and
the order parameter $\psi\to \xi^{-1}\psi$. Then in the
``canonical gauge'' the action assumes the form:
\be
S=\frac{\xi^{d-3}}{2u(2d)^{3/2}}\int d^{d+1}x\left[|\nabla
\psi|^2-|\psi|^2 +\half|\psi|^4\right],
\ee
which is identical to Eq. (\ref{gl}).
Second, to obtain the classical energy from (\ref{action0}) in the original
static gauge, we simply set time independent fields and
multiply by the energy unit $c/l$:
\be
{\mathcal E}=\frac{JN}{2ud} \int d^dx
\left[|\nabla\psi|^2-\left[r+(\mu l/c)^2\right]|\psi|^2+\half|\psi|^4\right]
\label{Energy0}
\ee
The average density offset form integer filling is then given by:
\be
\av{n-N}=-\frac{\partial {\mathcal E}}{\partial \mu}
=\frac{2\mu}{\a}|\psi|^2,
\label{n0}
\ee
which agrees with with Eq. (\ref{charge}) evaluated in the static
gauge, as it should.

Another note we should make is that for low filling factors
$N\sim 1$, the general form of the action remains the same, in the
vicinity of the Mott transition. However in that case, the
expressions for $\xi$, $c$, and the prefactor of the action are
more complicated.

\section{Derivation of the prefactor of the current decay in
a one-dimensional lattice}
\label{seca1}

Following a general theory developed in
Ref.~[\onlinecite{langer}], in the one dimensional case, we can
find the transition rates per site
$\Gamma_-$ and $\Gamma_+$. These can be written as:
\be
\Gamma_{\pm}={4\over \pi\tau}\sqrt{JN\over U}|\omega_0
\omega_{L/2}^0|\prod_{n\geq 1} {\omega_n^0\over \omega_n}\,\exp
\left(-{\Delta E_\pm\over T}\right),
\label{gen}
\ee
where $\tau$ is the phenomenological coupling to the bath degrees
of freedom, defined in terms of dissipative
Gross-Pitaveskii equations:
\be
\tau{\partial \psi_j\over \partial
t}=\psi_{j+1}+\psi_{j-1}-{U\over 2J}|\psi_j|^2\psi_j+\mathcal
L_j(t).
\ee
Here $\mathcal L_j(t)$ is the Langevin noise term; $\omega_j$ and
$\omega_j^0$ are the eigenvalues of the excitations of
Eq.~(\ref{eq7}) around the saddle-point and metastable states
respectively. Note that both states have one zero eigenvalue due
to global phase $U(1)$ symmetry and we ignore them. The saddle
point spectrum also has one imaginary eigenvalue ($\omega_0$)
corresponding to an unstable solution of the linearized equations of
motion. An extra prefactor $\omega_{L/2}^0$ comes because in the
saddle-point configuration the number of solutions with real
$\omega$ is smaller by one than in the metastable state. For
simplicity we assume $L$ to be even. Because of the absence of
continuous translational symmetry, there is no second zero
eigenvalue for the saddle point (compare with
Ref.~[\onlinecite{halperin}]). The eigenfunctions of small
fluctuations around the metastable state are plain-waves. The
corresponding spectrum thus reads:
\be
\omega_n=2\sqrt{2\cos p}\,\sin k_n/2,
\label{omega}
\ee
where $k_n=2\pi n/L$ is the momentum, $n=0,1,,.. L-1$ is an
integer and $L$ is the size of the chain.

The saddle point solution with a single phase slip has scattering
states and a bound state. The latter is described by the
eigen-function:
\be
\delta\phi_j=A\left\{\begin{array}{ll} \mathrm e^{-\kappa(j-1)}
&j\geq 1\\
-\mathrm e^{\kappa j} & j\leq 0.
\end{array}\right.
\label{kap}
\ee
Substituting (\ref{kap}) into the linearized version of
(\ref{eq7}) we find $\kappa=\ln 3$ and $\omega_0=4i\sqrt{2/3\,\cos
p}$, where we ignored a small discrepancy between $p$ and
$p^\prime$.

To find the scattering states we shall seek solutions in the
form:
\be
\delta\phi_j =A\mathrm e^{ikj}+B\mathrm e^{-ikj}
\ee
for $j=1,2,\dots L-1$ and $\delta\phi_0\equiv \delta\phi_N$, where
$\delta\phi_j$ is the small deviation from the saddle
point solution (\ref{eq8}). The system of secular equations
determining the wave vectors $k_n$ reads:
\begin{widetext}
\beq
&&A\left(1+\mathrm e^{ikL}-2\mathrm e^{ik}\right)+B\left(1+\mathrm
e^{-ikL}-2\mathrm
e^{-ik}\right)=0\\
&&A\mathrm e^{ik}\left(1+\mathrm e^{ikL}-2\mathrm
e^{ik(L-1)}\right)+B\mathrm e^{-ik}\left(1+\mathrm
e^{-ikL}-2\mathrm e^{-ik(L-1)}\right)=0.
\eeq
\end{widetext}
A nontrivial solution of the  above system exists for $k$
satisfying the following equation:
\be
\tan {kL\over 2}=2\tan {k\over 2}
\ee
Introducing the phase-shift $k_n=2\pi n/L+\delta_n/L$ we find
\be
\tan {\delta_n\over 2}=2\tan\left({\pi n\over L}+{\delta_n\over
2L}\right).
\ee
In the limit $L\to\infty$ we get the approximate solutions for the
scattering phaseshifts:
\be
\delta_n\approx 2\arctan\left( 2\tan{\pi n\over L}\right).
\ee
The energies of the scattering states are equal to
\be
\omega^\prime_n=2\sqrt{2\cos p^\prime}\sin\left({\pi n\over
L}+{\delta_n\over 2L}\right).
\ee
Now it is straightforward to find the ratio of the products of all
eigenvalues at the saddle point and the metastable state in the
limit $L\to \infty$:
\beq
&&\prod_n {\omega^\prime_{\mp}\over\omega_n}=3\,\mathrm e^{\pm\tan
p\,{\pi\mp2p\over 2}}.
\eeq
Substituting this into the general expression (\ref{gen}) we
derive (\ref{gam1}) and (\ref{gam2}).

\section{Gross-Pitaevskii dynamics of a lattice condensate in a parabolic
trap under slow change of the hopping amplitude.}
\label{seca3}

In the superfluid regime the most significant effect of the
optical trap on the motion of the system is the modification of
the effective mass of the particles. So, let us analyze a
condensate with a time dependent mass, moving in a parabolic trap.
Since we are considering a purely classical effect we can
use Gross-Pitaveskii equations (for simplicity we restrict the
analysis to one dimension:
\be
i{\partial \psi\over \partial t}=-{1\over 2m}{\partial^2\psi \over
\partial x^2}+{k\over 2}x^2\psi+U|\psi^2|\psi.
\label{1}
\ee
In the optical lattice the underlying equations are
\be
i{\partial \psi_j\over \partial
t}=-J(\psi_{j+1}+\psi_{j-1})+{k\over 2}j^2\psi+U|\psi^2|\psi.
\label{0},
\ee
In the weak tunneling regime the equations (\ref{1}) and (\ref{0})
are  equivalent provided that $m=1/(2J)$, the lattice constant is
equal to unity, and the phase gradient is small. If the
interaction strength is not too small, i.e. the condensate is in
the quantum-rotor limit $UN\gg J$, then the effect of quantum
pressure is negligible and instead of (\ref{1}) one can use
hydrodynamic equations of motion~\cite{brazhnyi}. The bosonic
field is then represented as:
\be
\psi(x,t)=\sqrt{\rho(x,t)} e^{i\int^x
p(x^\prime,t)dx^\prime}.
\ee
Keeping only the lowest orders of spatial derivatives of the
density $\rho$ instead of (\ref{1}) we obtain:
\beq
&&{\partial p\over\partial t}=-k x-{1\over m}p{\partial
p\over\partial x}-U {\partial \rho\over \partial x},\label{3}\\
&&{\partial\rho\over \partial t}=-{1\over m}{\partial\over\partial
x}(p\,\rho).
\label{4}
\eeq
The stationary solution of equations (\ref{3}) and (\ref{4})
yields an inverted parabola profile of the density:
\be
\rho_0(x)={\mu-{1\over 2}kx^2\over U},
\ee
where $\mu$ is the chemical potential.  Let us now assume that the
condensate undergoes small center of mass oscillations. They can
be excited by a small displacement of the trap minimum. Then it is
easy to check that one can seek a solution in the form:
\be
\rho(x,t)={\tilde\mu(t)-{k\over 2}x^2+f(t)x\over \lambda},\quad
f(t)=-{dp(t)\over dt},
\label{6}
\ee
with the initial conditions: $f(0)=k x_0$, $p(0)=0$,
$\mu(0)=\mu_0-{1\over 2}kx_0^2$, where $x_0$ is the initial
displacement. Substituting (\ref{6}) into (\ref{3}) and (\ref{4})
one finds:
\beq
&&f(t)=x_0k\cos\sqrt{k\over m}t,\quad
p(t)=-x_0\sqrt{km}\sin\sqrt{k\over m}t,\nonumber\\
&& \rho(x,t)={\mu-{1\over 2}k(x-{f(t)\over x_0})^2\over U}
\label{7}
\eeq
Note that the interaction $U$ never enters (\ref{7}) except for a
trivial prefactor. This is so because the excited mode is related
to the motion of the center of mass, while the shape of the
condensate cloud does not change in time. Moreover the existence
of such a center of mass or Galilean mode is the property of the
equation (\ref{1}) itself. Indeed, it is easy to check that if
$\psi_0(x,t)$ is the solution of (\ref{1}) then
\be
\psi(x,t)=\psi_0(x-s_0(t),t)\mathrm e^{ip(t) x}\mathrm e^{-{i\over
2} p(t)s_0(t)} \label{8}
\ee
is also a solution given that:
\be
{ds_0\over dt}={p(t)\over m},\; {dp(t)\over dt}=-k s_0(t).
\ee
Now let us assume that the mass increases in time. We will use
again the hydrodynamic equations (\ref{3}) and (\ref{4}), however
contrary to the stationary problem discussed above, the
hydrodynamic approximation is a crucial assumption for having the
center of mass mode independent of the interaction strength.
Strictly speaking, the Galilean invariance (\ref{8}) is valid only
if mass, curvature and interaction remain constant in time. In the
hydrodynamic regime the shape of the condensate does not depend on
the mass, therefore we may seek a solution in a form similar
to (\ref{6}):
\be
\rho(t)={\mu-{k\over 2}(x-{f(t)\over k})^2\over U},\; {dp\over
dt}=-f(t).
\ee
Substituting these formulas into (\ref{3}) and (\ref{4}) yields
\be
{d^2p\over dt^2}+{k\over m(t)}p=0
\ee
which is to be supplied with the initial conditions: $p(0)=0$,
$\dot p(0)=-x_0\,k$. In the adiabatic limit this equation has an
approximate solution:
\beq
&&p(t)\approx -x_0\sqrt{k} \left(m(t)m(0)\right)^{1\over
4}\sin\int_0^t \omega(\tau)d\tau,\label{14}\\
&&f(t)\approx x_0 k\left({m(0)\over m(t)}\right)^{1\over 4}
\cos\int_0^t \omega(\tau)d\tau.
\eeq
with $\omega(\tau)=\sqrt{k\over m(\tau)}$. The first of these
equations shows that as the mass increases (or equivalently the
tunneling constant decreases) the momentum also increases.


\begin{thebibliography}{99}
\bibitem{footote} We note that within Gutzwiller
approximation the size of the lattice is unimortant for finding
the position of the instability as long as the number of nearest
neighbors remains the same. The situation will change though, if
we are interested in the time evolution of the condensate after
the motion becomes unstable.
\end{thebibliography}
\end{document}